\documentclass[pre,showpacs,showkeys,preprintnumbers,amsmath,amssymb,superscriptaddress,twocolumn]{revtex4}
\usepackage{graphicx}
\usepackage{amsfonts}
\usepackage{url}
\usepackage{epstopdf}
\usepackage{color}
\usepackage{bm}
\usepackage[colorlinks=true,linkcolor=blue,citecolor=red]{hyperref}%

\def\L{\mathcal L}
\def\N{\mathcal N}
\def\var{{\rm var}}

\def\n{\bm{n}}
\def\s{\bm{s}}
\def\x{\bm{x}}
\def\X{\bm{X}}
\def\W{\bm{W}}

\def\pa{\partial\Omega}
\def\E{{\mathbb E}}
\def\I{{\mathbb I}}
\def\P{{\mathbb P}}
\def\R{{\mathbb R}}
\def\Z{{\mathbb Z}}
\def\T{{\mathcal T}}
\def\L{{\mathcal L}}

\def\M{{\mathcal M}}

\def\erfc{\mathrm{erfc}}
\def\erfcx{\mathrm{erfcx}}
\def\ctanh{\mathrm{ctanh}}

\begin{document}

\title{Probability distribution of the boundary local time \\ of reflected Brownian motion in Euclidean domains}

\author{Denis~S.~Grebenkov}
 \email{denis.grebenkov@polytechnique.edu}
\affiliation{
Laboratoire de Physique de la Mati\`{e}re Condens\'{e}e (UMR 7643), \\ 
CNRS -- Ecole Polytechnique, IP Paris, 91128 Palaiseau, France}

\date{\today}

\begin{abstract}
How long does a diffusing molecule spend in a close vicinity of a
confining boundary or a catalytic surface?  This quantity is
determined by the boundary local time, which plays thus a crucial role
in the description of various surface-mediated phenomena such as
heterogeneous catalysis, permeation through semi-permeable membranes,
or surface relaxation in nuclear magnetic resonance.  In this paper,
we obtain the probability distribution of the boundary local time in
terms of the spectral properties of the Dirichlet-to-Neumann operator.
We investigate the short-time and long-time asymptotic behaviors of
this random variable for both bounded and unbounded domains.  This
analysis provides complementary insights onto the dynamics of
diffusing molecules near partially reactive boundaries.
\end{abstract}





\maketitle

\section{Introduction}

Diffusion in confined media is common for many physical, chemical and
biological systems.  The presence of reflecting obstacles or reactive
surfaces drastically alters statistical properties of conventional
Brownian motion and controls diffusion-influenced phenomena such as
chemical reactions, surface relaxation or target search processes
\cite{Redner,Schuss,Metzler,Oshanin,Bouchaud90,Grebenkov07,Benichou14}.
A mathematical construction of such stochastic processes requires a
substantial modification of the underlying stochastic equation.  In
fact, a specific term has to be introduced into the stochastic
differential equation in order to ensure reflections and to prohibit
crossing a reflecting boundary.  In the simplest setting, the
reflected Brownian motion $\X_t$ in a given Euclidean domain $\Omega
\subset \R^d$ with a smooth enough boundary $\pa$ is constructed as
the solution of the stochastic Skorokhod equation
\cite{Ito,Freidlin,Anderson76,Brosamler76,Lions84,Saisho87,Hsu85,Williams87}:
\begin{equation}  \label{eq:Skorokhod}
d\X_t = \sigma \, d\W_t + \n(\X_t) \I_{\pa}(\X_t) d\ell_t,  \quad \X_0 = \x_0,
\end{equation}
where $\x_0 \in \bar{\Omega} = \Omega \cup \pa$ is a fixed starting
point, $\W_t$ is the standard $d$-dimensional Wiener process, $\sigma
> 0$ is the volatility, $\n(\x)$ is the normal unit vector at a
boundary point $\x$, which is perpendicular to the boundary at $\x$
and oriented outwards the domain $\Omega$, ${\mathbb I}_{\pa}(\x)$ is
the indicator function of the boundary (i.e., ${\mathbb I}_{\pa}(\x) =
1$ if $\x\in\pa$, and $0$ otherwise), and $\ell_t$ (with $\ell_0 = 0$)
is a nondecreasing process, which increases only when $\X_t\in\pa$,
known as the boundary local time.  The second term in
Eq. (\ref{eq:Skorokhod}), which is nonzero only on the boundary,
ensures that Brownian motion is reflected in the perpendicular
direction from the boundary.  The peculiar feature of this
construction is that the single Skorokhod equation determines
simultaneously two tightly related stochastic processes: $\X_t$ and
$\ell_t$.  Even though $\ell_t$ is called local time, it has units of
length, according to Eq. (\ref{eq:Skorokhod}).

In physics literature, the reflected Brownian motion is often
described without referring to the boundary local time $\ell_t$ by
using the heat kernel (also known as the propagator),
$G_0(\x,t|\x_0)$, which is the probability density of finding the
process $\X_t$ at time $t$ in a vicinity of $\x\in \bar{\Omega}$,
given that it was started from $\x_0 \in \bar{\Omega}$ at time $0$.
This heat kernel satisfies the diffusion equation
\begin{equation}
\partial_t G_0(\x,t|\x_0) = D \, \Delta_{\x} G_0(\x,t|\x_0) \qquad (\x\in\Omega),
\end{equation}
where $D = \sigma^2/2$ is the diffusion coefficient of reflected
Brownian motion, and $\Delta_{\x}$ is the Laplace operator acting on
$\x$.  This equation is completed by the initial condition
$G_0(\x,t=0|\x_0) = \delta(\x-\x_0)$ and {\it Neumann} boundary
condition:
\begin{equation}
\partial_n G_0(\x,t|\x_0)  = 0  \qquad (\x\in\pa),
\end{equation}
where $\partial_n = (\n(\x) \cdot \nabla)$ is the normal derivative
and $\delta(\x)$ is the Dirac distribution.  

\begin{figure}
\begin{center}
\includegraphics[width=50mm]{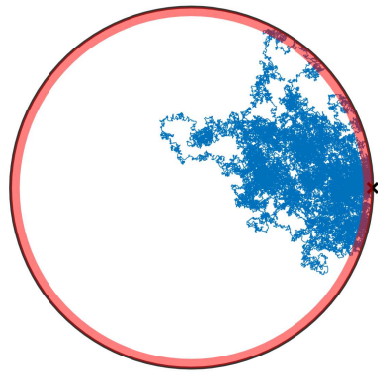}  
\end{center}
\caption{
A simulated reflected Brownian motion with diffusion coefficient $D$
inside a disk of radius $R$, up to time $t = R^2/D$.  Shadowed region
is a thin layer near the boundary of width $a/R = 0.05$.  The
residence time in this region, divided by $a$, is close to the
boundary local time $\ell_t$, see Eq. (\ref{eq:ell_res}).  Black cross
denotes the starting point of the trajectory.}
\label{fig:traj_disk}
\end{figure}

In turn, the boundary local time $\ell_t$ characterizes the behavior
of reflected Brownian motion $\X_t$ on the boundary $\pa$
(Fig. \ref{fig:traj_disk}).  As first described by P. L\'evy
\cite{Levy}, the boundary local time can be understood as the
renormalized residence time of $\X_t$ in a thin layer near the
boundary, $\pa_a = \{ \x \in \Omega~:~ |\x - \pa| < a\}$ up time $t$
\cite{Ito,Freidlin},
\begin{equation}  \label{eq:ell_res}
\ell_t = \lim\limits_{a \to 0} \frac{D}{a} \underbrace{\int\limits_0^t dt' \, \I_{\pa_a}(\X_{t'})}_{\textrm{residence time in}~\pa_a} .
\end{equation}
This relation highlights that the residence time in the boundary layer
$\pa_a$ vanishes in the limit $a\to 0$ when $\pa_a$ shrinks to the
boundary $\pa$.  This is not surprising given that the boundary $\pa$
has a lower dimension, $d-1$, as compared to the dimension $d$ of the
domain $\Omega$, and the residence time on the boundary is strictly
zero.  In turn, the rescaling of the residence time in $\pa_a$ by the
width $a$ of this layer yields a well-defined limit, namely, the
boundary local time.  Importantly, Eq. (\ref{eq:ell_res}) implies that
the residence time spent in a thin boundary layer $\pa_a$ can be
approximated as $a \ell_t/D$, as soon as $a$ is small enough.  The
boundary local time $\ell_t$ is thus the proper intrinsic
characteristics of reflected Brownian motion on the boundary, which is
independent of the layer width used.

The boundary local time $\ell_t$ is also related to the number
$\N_t^{a}$ of downcrossings of the boundary layer $\pa_a$ by reflected
Brownian motion up to time $t$, multiplied by $a$, in the limit $a\to
0$ \cite{Ito,Freidlin},
\begin{equation} \label{eq:ellt_N} 
\ell_t = \lim\limits_{a \to 0} a \, \N_t^{a} .
\end{equation}
The number of downcrossings can be mathematically defined by
introducing a sequence of interlacing hitting times $0 \leq
\delta_0^{(0)} < \delta_0^{(a)} < \delta_1^{(0)} < \delta_1^{(a)} <
\ldots$ as
\begin{subequations}
\begin{align}
\delta_n^{(0)} &= \inf\{ t > \delta_{n-1}^{(a)} ~:  ~\X_t \in \pa\},  \\
\delta_n^{(a)} &= \inf\{ t > \delta_n^{(0)} ~:  ~\X_t \in \Gamma_a\},
\end{align}
\end{subequations}
(with $\delta_{-1}^{(a)} = 0$), where $\Gamma_a = \{ \x\in\Omega ~:~
|\x - \pa| = a\}$.  Here, one records the first moment
$\delta_0^{(0)}$ when reflected Brownian motion hits the boundary
$\pa$, then the first moment $\delta_0^{(a)}$ of leaving the thin
layer $\pa_a$ through its inner boundary $\Gamma_a$, then the next
moment $\delta_1^{(0)}$ of hitting the boundary $\pa$, and so on.  In
this setting, the number of downcrossings of the thin layer $\pa_a$ up
to time $t$ (i.e., the number of excursions in the bulk) is the index
$n$ of the largest hitting time $\delta_n^{(0)}$, which is below $t$:
\begin{equation*}
\N_t^{a} = \sup\{ n > 0 ~:~ \delta_n^{(0)} < t\}.  
\end{equation*}
While the number of downcrossings diverges as $a\to 0$, its
renormalization by $a$ yields a well-defined limit $\ell_t$.
Conversely, the boundary local time divided by the layer width $a$,
$\ell_t/a$, is a proxy of the number of downcrossings of $\pa_a$, as
soon as $a$ is small enough.

One sees that the boundary local time characterizes the dynamics of a
diffusing particle near the boundary and thus plays a crucial role in
the description of various diffusion-mediated phenomena in cellular
biology, heterogeneous catalysis, nuclear magnetic resonance, etc.
\cite{Redner,Schuss,Metzler,Oshanin,Bouchaud90,Grebenkov07,Benichou14,Lauffenburger,Shoup82,Sapoval94,Sapoval02,Grebenkov05,Levitz06,Levitz08,Benichou10,Benichou10b,Rojo13,Bressloff13,Grebenkov19b}.
In these phenomena, a diffusing particle approaching the boundary can
change its state due to, e.g., permeation through a pore, chemical
reaction on a catalytic germ, or surface relaxation on a paramagnetic
impurity \cite{Grebenkov06,Grebenkov07a,Grebenkov09}.  As the related
interactions are typically short-ranged, the efficiency of such
surface mechanisms is directly related to the residence time of the
particle in a close vicinity of the boundary or, equivalently, to the
number of returns to that boundary, both being described by the
boundary local time.  In spite of its importance, the distribution of
the boundary local time in generic Euclidean domains and its
statistical properties are not well understood.  This is in contrast
to {\it point} local time processes whose properties were thoroughly
investigated, in particular, for Brownian motion and Bessel processes
(see \cite{Borodin,Takacs95,Randon18} and references therein).
Likewise, the residence (or occupation) time in a subset of a bounded
domain, which can be obtained by integrating the point local time over
the subset, was extensively studied for various diffusion processes
(see
\cite{Darling57,Agmon84,Berezhkovskii98,Majumdar05,Benichou05,Condamin05,Condamin07,Grebenkov07,Nguyen10}
and references therein).

In this paper, we provide a general description of the statistical
properties of the boundary local time $\ell_t$.  This description
relies on the spectral theory of diffusion-reaction processes with
heterogeneous surface reactivity developed in \cite{Grebenkov19}.  In
Sec. \ref{sec:theory}, we derive a spectral representation for the
probability density of the boundary local time $\ell_t$ in terms of
the eigenvalues and eigenfunctions of the Dirichlet-to-Neumann
operator.  We also establish the asymptotic behavior of the
probability density and of the moments of $\ell_t$.  In
Sec. \ref{sec:examples}, our general results are illustrated for
reflected Brownian motion inside and outside two archetypical
confinements: a disk and a ball.  Conclusions and perspectives of this
work are discussed in Sec. \ref{sec:conclusions}.

\section{General theory}
\label{sec:theory}

Our characterization of the boundary local time relies on two key
results: the construction of partially reflected Brownian motion
(Sec. \ref{sec:PRBM}) and the spectral representation of the
propagator via the Dirichlet-to-Neumann operator (Sec. \ref{sec:DtN}).

\subsection{Partially reflected Brownian motion}
\label{sec:PRBM}

In order to characterize the boundary local time $\ell_t$, we consider
a more general {\it partially reflected Brownian motion} (PRBM)
$\tilde \X_t$, whose heat kernel satisfies the diffusion
equation
\begin{equation}  \label{eq:G_diff}
\partial_t G_q(\x,t|\x_0) = D \, \Delta_{\x} G_q(\x,t|\x_0) \qquad (\x\in\Omega)
\end{equation}
for any $\x_0\in\bar{\Omega}$, subject to the initial condition
$G_q(\x,t=0|\x_0) = \delta(\x - \x_0)$ and the Robin (also known as
Fourier, radiation or third) boundary condition
\begin{equation}  \label{eq:Robin}
\partial_n G_q(\x,t|\x_0) + q \, G_q(\x,t|\x_0)  = 0  \quad (\x\in\pa)
\end{equation}
with a constant parameter
\begin{equation*}
q = \kappa/D \geq 0
\end{equation*}
(see \cite{Papanicolaou90,Bass08,Zhou16} for mathematical details and
references).  When the domain $\Omega$ is unbounded, one also needs to
impose a regularity condition at infinity: $G_q(\x,t|\x_0) \to 0$ as
$|\x|\to\infty$ (similar condition has to be imposed for the related
boundary value problems (\ref{eq:S_diff}, \ref{eq:tildeH_def},
\ref{eq:u_def}), see below).

The Robin boundary condition (\ref{eq:Robin}) appears in a large
variety of physical, chemical and biological applications
\cite{Collins49,Sano79,Sano81,Sapoval94,Benichou00,Sapoval02,Grebenkov05,Grebenkov06a,Bressloff08,Singer08,Grebenkov10a,Grebenkov10b,Rojo12,Grebenkov15}, 
as well as the effective boundary condition after homogenization
\cite{Zwanzig90,Berezhkovskii04,Berezhkovskii06,Muratov08,Dagdug16,Bernoff18b}
(see an overview in \cite{Grebenkov19b}).  The subscript $q$ allows us
to distinguish three types of boundary condition: Neumann ($q = 0$),
Robin ($0 < q < \infty$), and Dirichlet ($q = \infty$).  We note that
the notation $G_q(\x,t|\x_0)$ is different from that of
Refs. \cite{Grebenkov19,Grebenkov19b}, in which Neumann and Dirichlet
propagators were denoted as $G_{\kappa = 0}$ and $G_0$, respectively.

The partially reflected Brownian motion $\tilde \X_t$ can be defined
as reflected Brownian motion $\X_t$, which is stopped at the random
time $\T$ of reaction.  This stopping time is introduced by the
following reasoning (see
\cite{Grebenkov06,Grebenkov07a} for details).  At each arrival onto
the boundary, the particle either reacts with the probability $p =
1/(1 + D/(\kappa a))$, or resumes bulk diffusion from a distance $a$
above the boundary, with the probability $1-p$
\cite{Filoche99,Grebenkov03}.  Let $\hat{n}$ denote the random number
of failed attempts (reflections) before successful reaction.  As each
reaction attempt is independent from the others, one has $\P\{ \hat{n}
= n \} = p (1-p)^{n}$ (with $n = 0,1,2,\ldots$) and thus $\P\{ \hat{n}
\geq n \} = (1-p)^{n} \approx e^{-na \kappa/D}$ (for small $a$).
Since $\hat{n} \approx \ell_\T/a$ due to Eq. (\ref{eq:ellt_N}), we set
$\ell = na$ and thus get $\P\{ \ell_\T \geq \ell \} = e^{-\ell
\kappa/D}$ in the limit $a\to 0$; in other words, $\ell_\T$ obeys the
exponential distribution with the mean $D/\kappa$.  As the boundary
local time is a nondecreasing process, the event $\{\T > t\}$ is
identical to $\{ \ell_\T > \ell_t\}$:
\begin{equation} \label{eq:PP0}
\P_{\x_0}\{\T > t\} = \P_{\x_0}\{ \ell_\T > \ell_t \} \,.
\end{equation}
As a consequence, the stopping time $\T$ can be defined as the first
moment when the boundary local time $\ell_t$ exceeds a random
threshold $\hat{\ell}$ ($ = \ell_\T$):
\begin{equation} \label{eq:Tdef}
\T = \inf\{ t > 0 ~:~ \ell_t > \hat{\ell} \} ,
\end{equation}
where $\hat{\ell}$ is an independent exponential random variable with
the mean $D/\kappa$.  The independence follows from the fact that
$\ell_t$ is determined by the dynamics of the particle, whereas
$\hat{\ell} = \ell_\T$ is determined by the reactivity of the
boundary.

The cumulative distribution function of the stopping time $\T$,
$\P_{\x_0}\{\T \leq t\}$, is related to the survival probability of
the particle,
\begin{equation*}
S_q(t|\x_0) = \P_{\x_0}\{\T > t\} = 1 - \P_{\x_0}\{\T \leq t\}, 
\end{equation*}
which is obtained by integrating the propagator over the arrival point
$\x$:
\begin{equation}
S_q(t|\x_0) = \int\limits_\Omega d\x \, G_q(\x,t|\x_0).
\end{equation}
The survival probability also satisfies the diffusion equation
with Robin boundary condition:
\begin{subequations}  \label{eq:S_diff}
\begin{align}
\partial_t S_q(t|\x_0) & = D \Delta_{\x_0} S_q(t|\x_0) \quad (\x_0 \in \Omega) ,\\  \label{eq:S_Robin}
\partial_n S_q(t|\x_0) + q \, S_q(t|\x_0) & = 0  \quad (\x_0\in\pa),
\end{align}
\end{subequations}
with the initial condition $S_q(t=0|\x_0) = 1$, that follows from
Eqs. (\ref{eq:G_diff}, \ref{eq:Robin}) written in a backward form
\cite{Redner,Gardiner}.

Since $\ell_t$ and $\hat{\ell}$ are independent by construction, the
average over random realizations of $\hat{\ell}$ in Eq. (\ref{eq:PP0})
can be written as
\begin{equation}  \label{eq:S_rho}
S_q(t|\x_0) = \int\limits_0^\infty d\ell \, \underbrace{e^{-q \ell}}_{= \P\{ \hat{\ell} > \ell\} } \, \rho(\ell,t|\x_0)  \,,
\end{equation}
where $\rho(\ell,t|\x_0)$ is the probability density function (PDF) of
$\ell_t$ that we are looking for.  Even though Eq. (\ref{eq:S_rho})
fully determines $\rho(\ell,t|\x_0)$ via the inverse Laplace transform
with respect to $q$, the parameter $q$ is involved {\it implicitly} as
the coefficient in Robin boundary condition (\ref{eq:S_Robin}).  As a
consequence, even for simple domains like a disk or a ball, the above
relation accesses the PDF of the boundary local time $\ell_t$ only
numerically, and its practical implementation is time consuming.  In
the next section, we use a recently developed representation of the
survival probability in the basis of the Dirichlet-to-Neumann operator
\cite{Grebenkov19} in order to deduce a more explicit characterization
of the boundary local time.

\subsection{Spectral representation via Dirichlet-to-Neumann operator}
\label{sec:DtN}

The Laplace transform of Eq. (\ref{eq:S_rho}) with respect to time
$t$, denoted by tilde, reads
\begin{equation} \label{eq:Stilde_rho} 
\tilde{S}_q(p|\x_0) = \int\limits_0^\infty d\ell \, e^{-q \ell} \, \tilde{\rho}(\ell,p|\x_0)  .
\end{equation}
Writing the survival probability in terms of the PDF of the stopping
time $\T$, $H_q(t|\x_0)$,
\begin{equation} \label{eq:P_H}
\P_{\x_0}\{\T > t\} = 1 - \int\limits_0^t dt' \, H_q(t'|\x_0) ,
\end{equation}
one gets
\begin{equation}  \label{eq:auxil1} 
\frac{1-\tilde{H}_q(p|\x_0)}{p} = \int\limits_0^\infty d\ell \, e^{-q \ell} \, \tilde{\rho}(\ell,p|\x_0)  ,
\end{equation}
where 
\begin{equation} 
\tilde{H}_q(p|\x_0) = \E_{\x_0}\{ e^{-p\T}\} = \int\limits_0^\infty dt\, e^{-pt} \, H_q(t|\x_0)
\end{equation}
is the Laplace transform of $H_q(t|\x_0)$, and $\E_{\x_0}$ denotes the
expectation.  Applying the Laplace transform to Eqs. (\ref{eq:S_diff},
\ref{eq:P_H}), one easily shows that $\tilde{H}_q(p|\x_0)$ is the
solution of the following boundary value problem:
\begin{subequations}  \label{eq:tildeH_def}
\begin{align}
(p - D \Delta_{\x_0}) \tilde{H}_q(p|\x_0) &= 0 \quad (\x_0\in \Omega) ,\\
\biggl(\frac{1}{q} \partial_n  \tilde{H}_q(p|\x_0) + \tilde{H}_q(p|\x_0) \biggr) &= 1 \quad (\x_0\in \pa) .
\end{align}
\end{subequations} 
It is therefore convenient to express it in terms of the spectral
properties of the Dirichlet-to-Neumann operator $\M_p$
\cite{Grebenkov19}.

For a given function $f$ on the boundary $\pa$, the operator $\M_p$
associates another function on that boundary, $\M_p ~:~ f \mapsto g =
(\partial_n u)_{|\pa}$, where $u$ is the solution of the modified
Helmholtz equation subject to Dirichlet boundary condition:
\begin{subequations}  \label{eq:u_def}
\begin{eqnarray}  \label{eq:u_def1}
(p - D \Delta) u(\x) &=& 0 \quad (\x \in \Omega) , \\   \label{eq:u_def2}
u(\x) &=& f \quad (\x\in\pa).
\end{eqnarray}
\end{subequations}
In physical terms, if $f$ prescribes a concentration of particles
maintained on the boundary, then $\M_p f$ is proportional to the
steady-state diffusive flux density of these particles into the bulk
(with the bulk reaction rate $p$).  In mathematical terms, for a given
solution $u$ of the modified Helmholtz equation (\ref{eq:u_def1}), the
operator $\M_p$ maps the Dirichlet boundary condition, $u|_{\pa} = f$,
onto the equivalent Neumann boundary condition, $(\partial_n u)|_{\pa}
= g = \M_p f$.  Note that there is a family of operators $\M_p$
parameterized by $p \geq 0$.  For a smooth enough boundary $\pa$ (here
we skip conventional mathematical restrictions and rigorous
formulation of the involved functional spaces, see
\cite{Egorov,Jacob,Taylor,Marletta04,Arendt07,Arendt15,Hassell17,Girouard17}
for details), $\M_p$ is well-defined pseudo-differential self-adjoint
operator.

When the {\it boundary} is {\it bounded}, the spectrum of $\M_p$ is
discrete, i.e., there are infinitely many eigenpairs $\{\mu_n^{(p)},\,
v_n^{(p)}\}$, satisfying
\begin{equation}  \label{eq:M_eigen}
\M_p \, v_n^{(p)} = \mu_n^{(p)} \, v_n^{(p)}   \quad (n = 0,1,2,\ldots).
\end{equation}
The eigenvalues $\mu_n^{(p)}$ are nonnegative and growing to infinity
as $n\to\infty$, whereas the eigenfunctions $\{v_n^{(p)}\}$ form an
orthonormal complete basis of the space $L_2(\pa)$ of
square-integrable functions on $\pa$.  In order to rely on this
eigenbasis, we focus on bounded boundaries, whereas the confining
domain $\Omega$ can be bounded or not.  The limiting value of the
smallest eigenvalue $\mu_0^{(p)}$ as $p\to 0$ distinguishes two types
of diffusion: $\mu_0^{(0)} = 0$ for recurrent motion (diffusion in a
bounded domain in any dimension or diffusion in the exterior of a
compact set for $d = 2$) and $\mu_0^{(0)} > 0$ for transient motion
(diffusion in the exterior of a compact set for $d
\geq 3$).  Moreover, for diffusion in a bounded domain, the
corresponding eigenfunction is constant: $v_0^{(0)} = |\pa|^{-1/2}$.

On one hand, the action of the Dirichlet-to-Neumann operator can be
expressed by solving the boundary value problem (\ref{eq:u_def}) in a
standard way with the help of the Laplace-transformed propagator
$\tilde{G}_{\infty}(\x,p|\x_0)$ with Dirichlet boundary condition
($\kappa = \infty$):
\begin{align} 
& [\M_p f](\s_0) \\   \nonumber
& = \biggl(\partial_{n_0} \int\limits_{\pa} d\s \, 
\bigl(-D \partial_n \tilde{G}_{\infty}(\x,p|\x_0)\bigr)_{\x=\s} \, f(\s)\biggr)_{\x_0 = \s_0} .
\end{align}
On the other hand, the inverse of the Dirichlet-to-Neumann operator
for $p > 0$ can be expressed in terms of the Laplace-transformed
propagator $\tilde{G}_{0}(\x,p|\x_0)$ with Neumann boundary condition
($\kappa = 0$) \cite{Grebenkov19}:
\begin{equation}  \label{eq:Mp_inv}
D \tilde{G}_{0}(\s,p|\s_0) = \M_p^{-1} \delta(\s - \s_0)  \quad (\s,\s_0\in \pa)
\end{equation}
(note that $\M_0$ is not invertible for bounded domains).  We hasten
to outline a slight abuse of notation here and throughout the paper:
on the left-hand side of Eq. (\ref{eq:Mp_inv}), boundary points $\s$
and $\s_0$ are understood as points in $\R^d$ restricted to $\pa$; on
the right-hand side, boundary points $\s$ and $\s_0$ are understood as
points on a $(d-1)$-dimensional manifold $\pa$, on which the
Dirichlet-to-Neumann operator acts.  In particular, the
Laplace-transformed propagator has units of second $\cdot$
meter$^{-d}$, whereas the Dirac distribution has units of
meter$^{1-d}$.

Now we come back to the problem of finding the solution of
Eqs. (\ref{eq:tildeH_def}).  As shown in \cite{Grebenkov19},
$\tilde{H}_q(p|\x_0)$ admits the following spectral representation:
\begin{equation}  \label{eq:tildeH_x0}
\tilde{H}_q(p|\x_0) = \sum\limits_{n=0}^\infty \frac{V_n^{(p)}(\x_0) \int\nolimits_{\pa} d\s \, [v_n^{(p)}(\s)]^*}{1 + \mu_n^{(p)}/q} \,,
\end{equation}
where asterisk denotes complex conjugate, and
\begin{equation}  \label{eq:Vnp}
V_n^{(p)}(\x_0) = \int\limits_{\pa} d\s \, \tilde{j}_{\infty}(\s,p|\x_0) \, v_n^{(p)}(\s) ,
\end{equation}
with $\tilde{j}_{\infty}(\s,p|\x_0) = -D \bigl(\partial_n
\tilde{G}_{\infty}(\x,p|\x_0)\bigr)_{\x = \s}$ being the
Laplace transform of the probability flux density onto a perfectly
absorbing boundary (with Dirichlet boundary condition, $\kappa =
\infty$).

If the starting point $\x_0$ lies in the bulk $\Omega$, any trajectory
of the PRBM $\tilde \X_t$ can be split into two successive paths: from
$\x_0$ to a first hitting point $\s_0$ on the boundary, and from
$\s_0$ to a boundary point $\s$, at which the process is stopped.  The
stopping time $\T$ is thus the sum of two random durations of these
paths.  Along the first path, the boundary local time $\ell_t$ remains
zero and thus is not informative.  As first-passage times to a
boundary were thoroughly investigated in the past, it is convenient to
exclude this contribution from our analysis and to focus on the
second, much more complicated and less studied random variable.  For
this reason, we assume in the following that the starting point $\x_0$
lies on the boundary, i.e., $\x_0 = \s_0 \in \pa$.  In this case,
$\tilde{j}_{\infty}(\s,p|\s_0) = \delta(\s-\s_0)$ and thus
$V_n^{(p)}(\s_0) = v_n^{(p)}(\s_0)$ so that Eq. (\ref{eq:tildeH_x0})
is reduced to
\begin{equation}  \label{eq:tildeH}
\tilde{H}_q(p|\s_0) = \sum\limits_{n=0}^\infty \frac{\hat{v}_n^{(p)}(\s_0)}{1 + \mu_n^{(p)}/q} \,.
\end{equation}
where
\begin{equation}
\hat{v}_n^{(p)}(\s_0) = v_n^{(p)}(\s_0) \, \int\limits_{\pa} d\s \, [v_n^{(p)}(\s)]^* 
\end{equation}
are just the rescaled eigenfunctions $v_n^{(p)}(\s_0)$.
Once $\tilde{H}_q(p|\s_0)$ (or related quantity) is known for a starting
point $\s_0$ on the boundary, one can easily extend it to any starting
point $\x_0$ in the bulk using the relation:
\begin{equation}
\tilde{H}_q(p|\x_0) = \int\limits_{\pa} d\s_0 \, \tilde{j}_\infty(\s_0,p|\x_0) \, \tilde{H}_q(p|\s_0),
\end{equation}
which follows from Eqs. (\ref{eq:tildeH_x0}, \ref{eq:Vnp},
\ref{eq:tildeH}).  In particular, this relation applied to
Eq. (\ref{eq:Stilde_rho}) gives
\begin{align*}
& \underbrace{\int\limits_0^\infty d\ell \, e^{-q\ell} \, \tilde{\rho}(\ell,p|\x_0)}_{ = \tilde{S}_q(p|\x_0)} = \tilde{S}_\infty(p|\x_0)  \\
& + \int\limits_{\pa} d\s_0 \, \tilde{j}_\infty(\s_0,p|\x_0) 
\underbrace{\int\limits_0^\infty d\ell \, e^{-q\ell} \, \tilde{\rho}(\ell,p|\s_0)}_{= \tilde{S}_q(p|\s_0)} ,
\end{align*}
from which the inverse Laplace transform with respect to $q$ yields
\begin{equation}
\tilde{\rho}(\ell,p|\x_0) = \tilde{S}_\infty(p|\x_0) \, \delta(\ell) 
+ \int\limits_{\pa} d\s_0 \, \tilde{j}_\infty(\s_0,p|\x_0) \, \tilde{\rho}(\ell,p|\s_0),
\end{equation}
whereas the inverse Laplace transform with respect to $p$ leads to
\begin{align}  \label{eq:rho_x0}
\rho(\ell,t|\x_0) & = S_\infty(t|\x_0) \, \delta(\ell)  \\   \nonumber
& + \int\limits_{\pa} d\s_0 \int\limits_0^t dt'\, j_\infty(\s_0,t'|\x_0) \, \rho(\ell,t-t'|\s_0).
\end{align}
This relation has a simple probabilistic interpretation.  When the
particle starts from a bulk point $\x_0\in \Omega$, the boundary local
time remains zero until the first arrival onto the boundary.  As a
consequence, the probability distribution of $\ell_t$ has an atom at
$\ell = 0$, i.e., $\ell_t$ is zero with a finite probability, which is
equal to the survival probability $S_\infty(t|\x_0)$ (the first term).
In turn, the positive values of $\ell_t$ are given by the convolution
of the probability density of arriving at $\s_0$ at time $t'$ with the
probability density of getting $\ell$ within the remaining time $t-t'$
from the starting point $\s_0$ (the second term).  As
Eq. (\ref{eq:rho_x0}) expresses the probability density
$\rho(\ell,t|\x_0)$ for any bulk point $\x_0$ in terms of
$\rho(\ell,t|\s_0)$ for a boundary point $\s_0$, we focus on the
latter quantity in the reminder of the paper.

The completeness of eigenfunctions $v_n^{(p)}$ implies the identity
\begin{equation}  \label{eq:completeness}
\sum\limits_{n=0}^\infty \hat{v}_n^{(p)}(\s_0) = 1.
\end{equation}
Using this representation of $1$, one can rewrite
Eq. (\ref{eq:auxil1}) as
\begin{equation}
\frac{1}{p} \sum\limits_{n=0}^\infty \hat{v}_n^{(p)}(\s_0) \frac{\mu_n^{(p)}}{\mu_n^{(p)} + q}   
 = \int\limits_0^\infty d\ell \, e^{-q \ell} \, \tilde{\rho}(\ell,p|\s_0)  \,,
\end{equation}
from which
\begin{equation}  \label{eq:rho_DtN}
\tilde{\rho}(\ell,p|\s_0) = \frac{1}{p} \sum\limits_{n=0}^\infty \hat{v}_n^{(p)}(\s_0) \, \mu_n^{(p)} \, e^{-\mu_n^{(p)} \ell} \,.
\end{equation}
The inverse Laplace transform with respect to $p$ yields the PDF
$\rho(\ell,t|\s_0)$ of the boundary local time $\ell_t$:
\begin{equation}  \label{eq:rho_DtN_inv}
\rho(\ell,t|\s_0) = \L^{-1}_t \biggl\{\frac{1}{p} \sum\limits_{n=0}^\infty \hat{v}_n^{(p)}(\s_0) \, \mu_n^{(p)} \, e^{-\mu_n^{(p)} \ell} \biggr\} \,.
\end{equation}
Since
\begin{equation}  \label{eq:rho_P}
\rho(\ell,t|\s_0) = - \frac{\partial \P_{\s_0}\{ \ell_t > \ell\}}{\partial \ell} \,,
\end{equation}
the integral of Eq. (\ref{eq:rho_DtN}) from $\ell$ to infinity gives
\begin{equation}  \label{eq:P_DtN}
\int\limits_0^\infty dt \, e^{-pt} \, \P_{\s_0}\{ \ell_t > \ell\} = 
\frac{1}{p} \sum\limits_{n=0}^\infty \hat{v}_n^{(p)}(\s_0)  \,  e^{-\mu_n^{(p)} \ell} \,,
\end{equation}
and thus
\begin{equation}  \label{eq:P_DtN_inv}
\P_{\s_0}\{ \ell_t > \ell\} = \L^{-1}_t \left\{
\frac{1}{p} \sum\limits_{n=0}^\infty \hat{v}_n^{(p)}(\s_0)  \,  e^{-\mu_n^{(p)} \ell} \right\} \,.
\end{equation}
Either of Eqs. (\ref{eq:rho_DtN}, \ref{eq:P_DtN}) fully determines the
distribution of the boundary local time $\ell_t$.  These are the main
results of the paper.  While we treated the boundary as reactive to
define the stopping time $\T$ and to perform the above derivation, the
final results (\ref{eq:rho_DtN}, \ref{eq:P_DtN}) do not depend on the
reactivity $\kappa$.  Indeed, these relations determine the boundary
local time and thus characterize the dynamics near reflecting
boundary, which is disentangled from eventual surface reactions.
Note that Eq. (\ref{eq:completeness}) implies $\P_{\s_0}\{
\ell_t > 0\} = 1$ that is equivalent to the normalization of the
probability density $\rho(\ell,t|\s_0)$.

The relation (\ref{eq:rho_DtN}) also determines the positive moments
of the boundary local time in the Laplace domain:
\begin{equation}  \label{eq:mean_DtN}
\int\limits_0^\infty dt \, e^{-pt} \, \E_{\s_0}\{ \ell_t^k \} = 
\frac{k!}{p} \sum\limits_{n=0}^\infty \frac{\hat{v}_n^{(p)}(\s_0)}{[\mu_n^{(p)}]^{k}}   \,.
\end{equation}

\subsection{Short-time behavior}

For $k = 1$, the sum in the right-hand side of Eq. (\ref{eq:mean_DtN})
can be seen as the spectral representation of the inverse of the
Dirichlet-to-Neumann operator, $\M_p^{-1}$, which is equal to $D
\tilde{G}_{0}(\s,p|\s_0)$ according to Eq. (\ref{eq:Mp_inv}).
As a consequence, the Laplace transform can be inverted to get
\begin{equation}  \label{eq:mean_DtN1}
\E_{\s_0}\{ \ell_t \} = \int\limits_0^t dt' \, \int\limits_{\pa} d\s \, D G_{0}(\s,t'|\s_0) \,.
\end{equation}
This representation also follows directly from the general formula for
the residence time and its limiting form in Eq. (\ref{eq:ell_res}).
In the short-time limit, the propagator can be locally approximated by
that near a reflecting hyperplane,
\begin{equation}
G_{0}(\s,t|\s_0) \simeq \frac{\exp\bigl(-|\s-\s_0|^2/(4Dt)\bigr)}{(4\pi Dt)^{(d-1)/2}} \, \frac{1}{\sqrt{\pi Dt}} \,,
\end{equation}
where the second factor accounts for the orthogonal direction.
Integrating this function over $\s \in \R^{d-1}$, one gets from
Eq. (\ref{eq:mean_DtN1}):
\begin{equation}  \label{eq:mean_DtN1_t0}
\E_{\s_0}\{ \ell_t \} \simeq 2\sqrt{D t}/\sqrt{\pi}  \qquad (t\to 0).
\end{equation}
Here, the short-time behavior does not depend on the starting point
$\s_0$ because the boundary locally looks flat as $t\to 0$.  This
asymptotic behavior agrees with the upper bound provided in
\cite{Hsu85}.  Qualitatively, this universal asymptotic behavior
can be rationalized as following.  At short times, the particle moves
away from the boundary by a distance of the order of $\sqrt{Dt}$,
i.e., the typical available volume is $(\sqrt{Dt})^d$ (here, we omit
eventual numerical prefactors), in which the residence time is close
to $t$.  The mean residence time in a thin boundary layer of width $a$
and of lateral radius $\sqrt{Dt}$, whose volume is of the order $a
(\sqrt{Dt})^{d-1}$, is the total residence time (close to $t$),
multiplied by the ratio of these volumes: $t \,
a(\sqrt{Dt})^{d-1}/(\sqrt{Dt})^{d}$.  According to
Eq. (\ref{eq:ell_res}), the mean boundary local time is then
$\sqrt{Dt}$, up to the numerical constant (given in
Eq. (\ref{eq:mean_DtN1_t0})).

\subsection{Long-time behavior}
\label{sec:long}

To study the long-time behavior, we distinguish three cases.

\subsubsection*{Diffusion in a bounded domain}

Diffusion in a bounded domain is recurrent in any space $\R^d$ so that
$\mu_0^{(p)} \to 0$ as $p\to 0$.  More precisely, one has (see
Appendix \ref{sec:mu0_p})
\begin{equation}  \label{eq:mu0_p0}
\mu_0^{(p)}\simeq \frac{p}{D} \, \frac{|\Omega|}{|\pa|}  \qquad (p\to 0)
\end{equation}
(here $|A|$ is the Lebesgue measure of $A$), while $v_0^{(p)} \to
v_0^{(0)} = |\pa|^{-1/2}$ so that the orthogonality of eigenfunctions
$\{v_n^{(0)}\}$ simplifies Eq. (\ref{eq:mean_DtN}) and yields
\footnote{
At a first look, our asymptotic result (\ref{eq:mean_DtN_asympt})
disagrees with the upper bound provided on p. 446 of
\cite{Hsu85}.  However, this upper bound was based on the standard
{\it short-time} upper bound by $K/\sqrt{t}$ of the transition density
function (i.e., the propagator) and thus is only applicable at short
times.  Clearly, the upper bound $K/\sqrt{t}$ does not hold at long
times, at which the propagator $G_0(\x,t|\x_0)$ approaches a constant,
$1/|\Omega|$.  In other words, the statement on p. 446 of \cite{Hsu85}
should be amended as: there are positive constants $t_0$ and $K_n$
such that $\sup_{\x\in\bar{\Omega}} \E_{\x}\{\ell^n(t)\} \leq K_n
t^{n/2}$ for all $t \leq t_0$.}
%
\begin{equation}  \label{eq:mean_DtN_asympt}
\E_{\s_0}\{ \ell_t^k \} \simeq (Dt |\pa|/|\Omega|)^k  \qquad (t\to\infty).
\end{equation}
As expected, these moments grow up to infinity as $t\to\infty$, and
the long-time asymptotic behavior does not depend on the starting
point $\s_0$.  In particular, the linear growth of the mean boundary
local time with $t$ has a simple explanation: at long times, the
particle is uniformly distributed in the bounded domain and thus
spends in a thin boundary layer $\pa_a$ a fraction of time, which is
proportional to the volume of $\pa_a$ divided by the volume of the
domain $\Omega$.  In other words, the mean residence time in $\pa_a$
is approximately $t |\pa_a|/|\Omega| \approx t a|\pa|/|\Omega|$, from
which Eq. (\ref{eq:ell_res}) yields $\E_{\s_0}\{ \ell_t \} \simeq Dt
|\pa|/|\Omega|$, in agreement with Eq. (\ref{eq:mean_DtN_asympt}).

In \cite{Grebenkov07a}, a much stronger property was established: all
the cumulant moments of $\ell_t$ grow linearly with time $t$.  As a
consequence, the distribution of the boundary local time is
asymptotically close to a Gaussian distribution in the limit $t\to
\infty$:
\begin{equation}  \label{eq:rho_Gauss}
\rho(\ell,t|\s_0) \simeq \frac{\exp\bigl(- \frac{(\ell - Dt |\pa|/|\Omega|)^2}{2b_{2,1} t}\bigr)}{\sqrt{2\pi b_{2,1} t}} \quad (t\to\infty),
\end{equation}
where the constant $b_{2,1}$ was formally computed in
\cite{Grebenkov07a}.  In Appendix \ref{sec:b21}, we express this
constant in terms of the second derivative of the smallest eigenvalue
$\mu_0^{(p)}$ with respect to $p$ (evaluated at $p = 0$):
\begin{equation}
b_{2,1} = - \biggl(\frac{D |\pa|}{|\Omega|}\biggr)^3 \lim\limits_{p\to 0} \frac{d^2 \mu_0^{(p)}}{dp^2} \,.
\end{equation}

\subsubsection*{Diffusion in the exterior of a compact planar set}

When $\Omega$ is the exterior of a compact planar set, diffusion is
still recurrent, and $\mu_0^{(p)} \to 0$ as $p\to 0$.  However, the
approach to zero is much slower than in Eq. (\ref{eq:mean_DtN}).  In
this setting, the mean boundary local time also grows up to infinity
but much slower (see Sec. \ref{sec:DiskE} for an example in the
exterior of a disk).

\subsubsection*{Diffusion in the exterior of a compact set in higher dimensions}

When $\Omega$ is the exterior of a compact set in $\R^d$ with $d \geq
3$, one has $\mu_0^{(p)} \to \mu_0^{(0)} > 0$ as $p\to 0$, diffusion
is transient, i.e., the particle will ultimately escape to infinity
and never return.  As a consequence, Eq. (\ref{eq:P_DtN}) implies
\begin{equation}  \label{eq:P_DtN_asympt}
\P_{\s_0}\{ \ell_t > \ell\} \to \P_{\s_0}\{ \ell_\infty > \ell\}  \qquad (t\to \infty),
\end{equation}
with
\begin{equation}
\P_{\s_0}\{ \ell_\infty > \ell\} = 
\sum\limits_{n=0}^\infty \hat{v}_n^{(0)}(\s_0) \, e^{-\mu_n^{(0)} \ell} \,.
\end{equation}
In other words, the boundary local time reaches its steady-state limit
$\ell_\infty$ determined by the above distribution and the following
moments:
\begin{equation}  \label{eq:mean_DtN_tau_tinf}
\E_{\s_0}\{ \ell_\infty^k \} = k! \sum\limits_{n=0}^\infty \frac{\hat{v}_n^{(0)}(\s_0)}{[\mu_n^{(0)}]^{k}}   \,.
\end{equation}
We emphasize that $v_n^{(0)}(\s)$ is not in general constant for
exterior diffusion so that all eigenmodes can contribute.

\subsection{A probabilistic interpretation}

Introducing an independent exponentially distributed random stopping
time $\tau$, defined by the rate $p$ as $\P\{\tau > t\} = e^{-pt}$,
one can multiply the left-hand side of Eq. (\ref{eq:P_DtN}) by $p$ and
interpret it as the average over the exponential stopping time $\tau$
(with the probability density $p\, e^{-pt}$)
\begin{equation}  \label{eq:PP}
\P_{\s_0} \{ \ell_\tau > \ell\} = \int\limits_0^\infty  dt \, p \, e^{-pt} \, \P_{\s_0} \{ \ell_t > \ell\} .
\end{equation}
In other words, we get explicitly the probability law for the boundary
local time $\ell_\tau$ stopped at an exponentially distributed time
$\tau$:
\begin{equation}  \label{eq:P_DtN_2}
\P_{\s_0}\{ \ell_\tau > \ell\} = 
\sum\limits_{n=0}^\infty \hat{v}_n^{(p)}(\s_0) \, e^{-\mu_n^{(p)} \ell} \,.
\end{equation}
Similarly, Eq. (\ref{eq:mean_DtN}) yields the moments of the boundary
local time stopped at $\tau$:
\begin{equation}  \label{eq:mean_DtN_tau}
\E_{\s_0}\{ \ell_\tau^k \} = k! \sum\limits_{n=0}^\infty \frac{\hat{v}_n^{(p)}(\s_0)}{[\mu_n^{(p)}]^{k}}   \,.
\end{equation}

The probabilistic interpretation of $\ell_\tau$ is rather
straightforward in terms of ``mortal walkers''
\cite{Yuste13,Meerson15,Grebenkov17d}.  In fact, one can consider a
particle that diffuses in a reactive bulk and can spontaneously
disappear with the rate $p$.  In this setting, $\tau$ is the random
lifetime of such a mortal walker.

\section{Examples}
\label{sec:examples}

In this section, we illustrate the properties of the boundary local
time with five examples, for which the eigenbasis of the
Dirichlet-to-Neumann operator is known explicitly.  The probability
density function $\rho(\ell,t|\s_0)$ is then obtained by the numerical
inversion of the Laplace transform in Eq. (\ref{eq:rho_DtN_inv}) using
the Talbot algorithm.  The accuracy of this numerical computation was
validated by Monte Carlo simulations presented in Appendix
\ref{sec:validation}.

\subsection{Half-space}

The simplest setting for the analysis of the boundary local time
$\ell_t$ is the half-space $\R^d_+$.  Formally, one would need to
consider the Dirichlet-to-Neumann operator on a hyperplane which is
the boundary of this domain, and thus to deal with continuous
spectrum.  However, the translational invariance of the half-space
implies that the lateral motion along the hyperplane is independent of
the transverse motion, which thus fully determines $\ell_t$.  In other
words, the boundary local time on a hyperplane is identical to that on
the endpoint of the positive half-line $\R_+ = (0,+\infty)$ with
reflections at $0$.  The latter is twice the local time of Brownian
motion at zero that was thoroughly investigated starting from the
seminal works by P. L\'evy \cite{Levy} (see also \cite{Takacs95}).

For illustrative purposes, we rederive its distribution from our
general approach.  The derivation is particularly simple because the
boundary of the half-line is just a single point so that the
Dirichlet-to-Neumann operator acts on a one-dimensional space of
functions.  In fact, a general solution of the modified Helmholtz
equation (\ref{eq:u_def1}) is $u(x) = f \, \exp(-x\sqrt{p/D})$ with a
constant $f$ set by the boundary condition (\ref{eq:u_def2}), while
its normal derivative at zero is $f \sqrt{p/D}$.  The action of $\M_p$
is thus the multiplication of a function at the boundary, namely, a
constant $f$, by $\sqrt{p/D}$.  There exists a single eigenvalue of
$\M_p$, $\mu_0^{(p)} = \sqrt{p/D}$, with the corresponding
eigenfunction $v_0^{(p)} = 1$.  According to
Eq. (\ref{eq:rho_DtN_inv}), the probability density of the boundary
local time is then
\begin{equation}
\rho(\ell,t) = \L^{-1}_t \left\{ \frac{\sqrt{p/D}}{p} \, e^{-\ell \sqrt{p/D}} \right\} 
= \frac{\exp\bigl(- \frac{\ell^2}{4Dt}\bigr)}{\sqrt{\pi Dt}} \,.
\end{equation}
A similar computation can be undertaken for an interval.

\subsection{Interior of a disk}
\label{sec:Dinterior}

We then study the local time on the boundary $\pa$ of a disk of radius
$R$, $\Omega = \{\x\in \R^2 ~:~ |\x| < R\}$.  Even though the
eigenmodes of the Dirichlet-to-Neumann operator $\M_p$ are well known
for this domain, we rederive them to illustrate the method.  For this
purpose, one needs to solve the Dirichlet boundary value problem
(\ref{eq:u_def}).  Due to the rotational symmetry of the domain
$\Omega$, one can search a general solution of the modified Helmholtz
equation (\ref{eq:u_def1}) in polar coordinates $(r,\theta)$ in the
form
\begin{equation}  \label{eq:u_Dint}
u(r,\theta) = \sum\limits_{n = -\infty}^\infty c_n \, I_n(r \sqrt{p/D}) \, e^{in\theta} ,
\end{equation}
where $I_n(z)$ are the modified Bessel functions of the first kind,
and the coefficients $c_n$ are fixed by the Dirichlet condition
(\ref{eq:u_def2}) with a given function $f$:
\begin{equation}
c_n = \frac{1}{I_n(R\sqrt{p/D})} \int\limits_{0}^{2\pi} \frac{d\theta}{2\pi} f(\theta) \, e^{-in\theta} .
\end{equation}
As the normal derivative acts only on the radial coordinate,
$\partial_n = \partial_r$, the action of $\M_p$ onto $f$ reads
\begin{align*}
\M_p f & = \bigl(\partial_n u(r,\theta)\bigr)_{|\pa} \\
& = \sum\limits_{n = -\infty}^\infty \frac{\sqrt{p/D}\, I'_n(R \sqrt{p/D})}{I_n(R\sqrt{p/D})} \, e^{in\theta} 
\int\limits_{0}^{2\pi} \frac{d\theta}{2\pi} f(\theta) \, e^{-in\theta},
\end{align*}
where prime denotes the derivative with respect to the argument.
Setting $f(\theta) = e^{in\theta}$, one has 
\begin{equation}
\M_p e^{in\theta} = \frac{\sqrt{p/D}\, I'_n(R \sqrt{p/D})}{I_n(R\sqrt{p/D})} \, e^{in\theta}  \,,
\end{equation}
i.e., $e^{in\theta}$ is an eigenfunction of $\M_p$ for any $n\in \Z$,
whereas
\begin{equation}  \label{eq:mu_diskI}
\mu_n^{(p)} = \sqrt{p/D} \, \frac{I'_n(R \sqrt{p/D})}{I_n(R\sqrt{p/D})}  
\end{equation}
is the corresponding eigenvalue.  We emphasize that the form of the
eigenfunctions is a direct consequence of the rotational symmetry of
the domain.  For coherence with the general description in
Sec. \ref{sec:theory}, we substitute the angular coordinate $\theta$
by the curvilinear coordinate $s/R$, with $s$ ranging from $0$ to
$2\pi R$ along the circular boundary $\pa$,
\begin{equation}
v_n^{(p)}(s) = \frac{e^{ins/R}}{\sqrt{2\pi R}}  \qquad (n \in \Z),
\end{equation}
in which the $L_2(\pa)$-normalization is also incorporated.  In this
particular example, the eigenfunctions do not depend on $p$, whereas
the eigenvalues are twice degenerate, except for $n = 0$.  Here, the
index $n$ runs over all integer numbers for convenience of
enumeration.

The orthogonality of the harmonics $\{e^{ins/R}\}$ to a constant
implies that only the term with $n= 0$ survives in
Eqs. (\ref{eq:rho_DtN}, \ref{eq:P_DtN}), yielding
\begin{equation}  \label{eq:P_DtN_disk}
\P_{\s_0}\{ \ell_t > \ell\} = \L^{-1}_t \left\{\frac{1}{p} \exp\left(- \ell \sqrt{p/D} \frac{I_1(R \sqrt{p/D})}{I_0(R\sqrt{p/D})} \right)\right\} ,
\end{equation}
from which $\rho(\ell,t)$ is found via Eq. (\ref{eq:rho_P}).  As
expected, this result does not depend on the starting point $\s_0$ on
the circle.  The mean boundary local time from Eq. (\ref{eq:mean_DtN})
reads
\begin{equation}  \label{eq:mean_DtN_disk}
\E\{ \ell_t \} = \L^{-1}_t \left\{ \frac{1}{p} \,  \frac{I_0(R\sqrt{p/D})}{\sqrt{p/D} \, I_1(R \sqrt{p/D})} \right\} \,.
\end{equation}
From this expression, one easily retrieves the short-time and
long-time asymptotic behaviors: $\E\{ \ell_t \} \simeq
2\sqrt{Dt}/\sqrt{\pi}$ as $t\to 0$ and $\E\{ \ell_t \} \simeq 2Dt/R$
as $t\to \infty$, in agreement with Eqs. (\ref{eq:mean_DtN1_t0},
\ref{eq:mean_DtN_asympt}).  We emphasize that
Eqs. (\ref{eq:P_DtN_disk}, \ref{eq:mean_DtN_disk}) also characterize
the boundary local time of reflected Brownian motion inside a cylinder
of radius $R$ (given that displacements along the cylinder axis do not
affect the boundary local time).  In particular, $\ell_t$ determines
the residence time in a thin cylindrical layer and the number of
returns to this layer.

\begin{figure}
\begin{center}
\includegraphics[width=88mm]{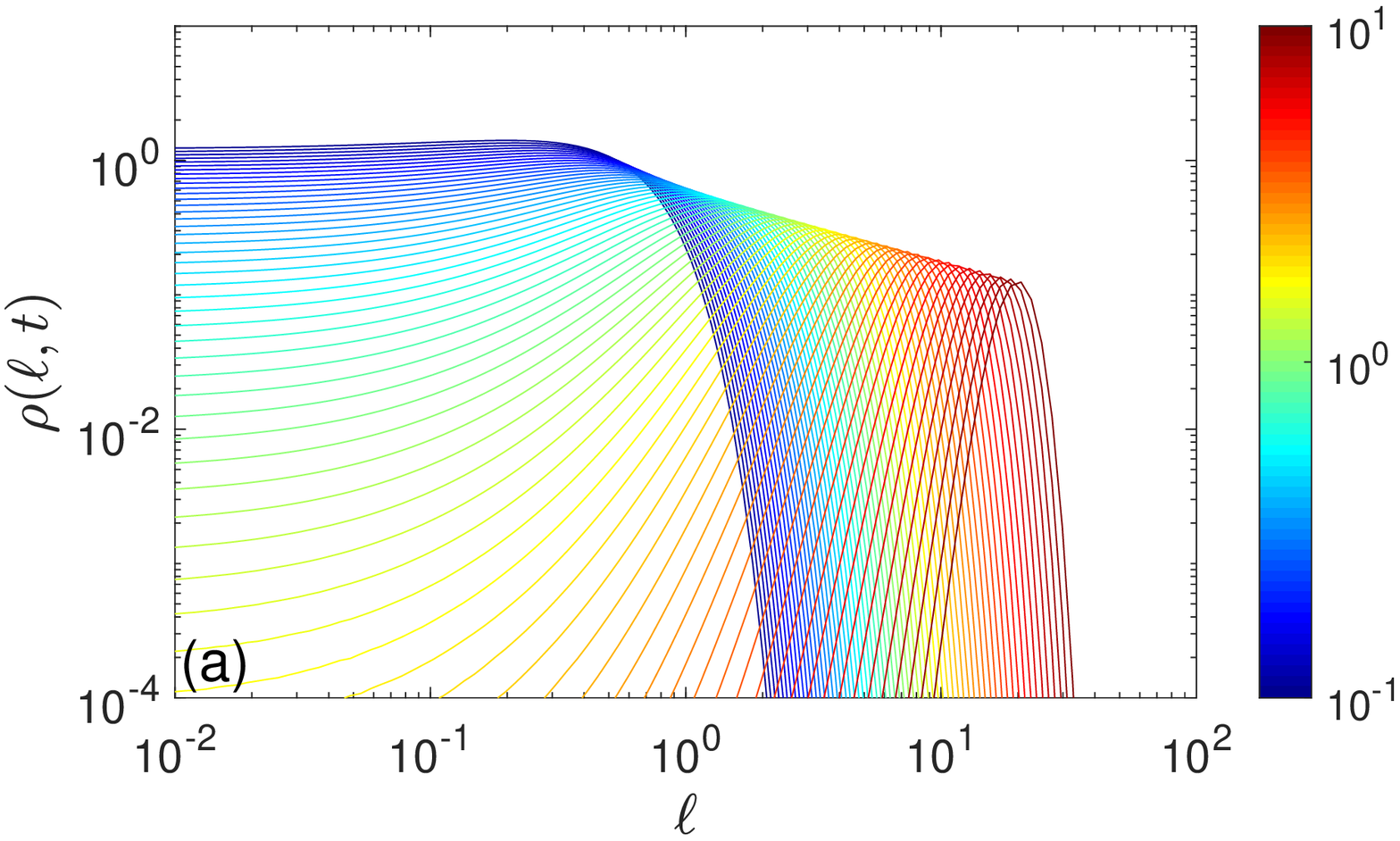} 
\includegraphics[width=88mm]{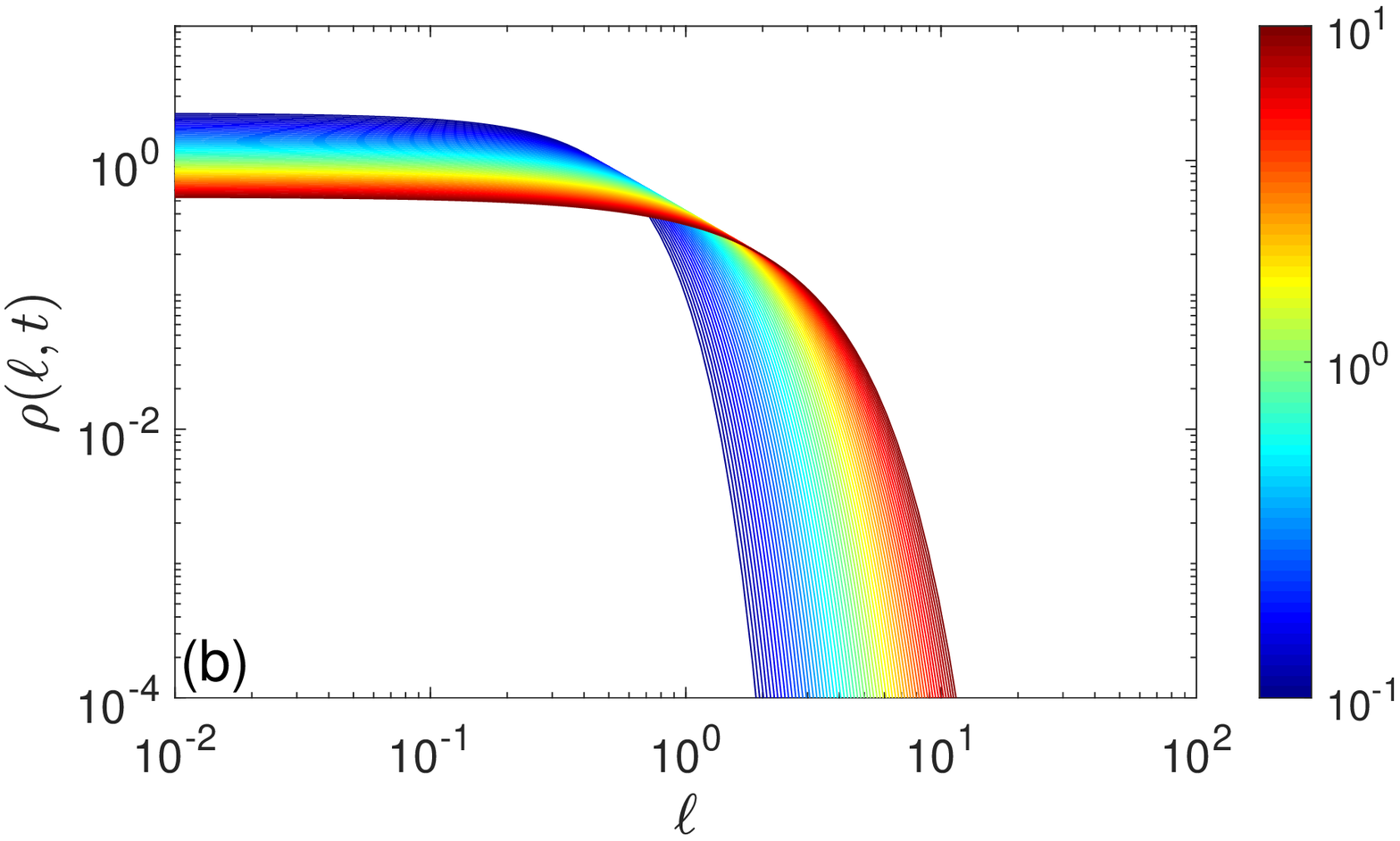} 
\end{center}
\caption{
Probability density function $\rho(\ell,t)$ of the boundary local time
$\ell_t$ for a disk of radius $R = 1$, with $D = 1$ and $t$ taking 64
logarithmically spaced values from $10^{-1}$ (dark blue) to $10^1$
(dark red).  {\bf (a)} diffusion inside the disk; {\bf (b)} diffusion
outside the disk.}
\label{fig:rho_disk}
\end{figure}

Figure \ref{fig:rho_disk}a shows the probability density function
$\rho(\ell,t)$ for different times $t$.  One can notice that
$\rho(\ell,t)$ exhibits a maximum, which is progressively shifted
toward larger $\ell$ with time.  At short times (blue curves), the PDF
is flat at small $\ell$, and then rapidly drops at large $\ell$.  As
time $t$ increases, the shape of the PDF transforms and becomes more
localized near the mean boundary local time.  At long times (red
curves), the PDF is getting close to a Gaussian distribution
(\ref{eq:rho_Gauss}), with the linearly growing mean and variance, as
discussed in Sec. \ref{sec:long}.

\subsection{Exterior of a disk}
\label{sec:DiskE}

For the exterior of a disk of radius $R$, $\Omega = \{\x\in\R^2 ~:~
|\x| > R\}$, the eigenfunctions of the Dirichlet-to-Neumann operator
remain unchanged (as a consequence of the preserved rotational
symmetry), whereas the eigenvalues are
\begin{equation}  \label{eq:mu_diskE}
\mu_n^{(p)} = -\sqrt{p/D} \frac{K'_n(R \sqrt{p/D})}{K_n(R\sqrt{p/D})}  \qquad (n\in \Z) \,,
\end{equation}
where $K_n(z)$ are the modified Bessel functions of the second kind.
Indeed, one can repeat the derivation from Sec. \ref{sec:Dinterior} by
replacing $I_n(r\sqrt{p/D})$ in Eq. (\ref{eq:u_Dint}) by
$K_n(r\sqrt{p/D})$, which vanish as $r\to\infty$, and using
$\partial_n = - \partial_r$, which results in the negative sign in
Eq. (\ref{eq:mu_diskE}).

As previously, the orthogonality of eigenfunctions
reduces Eq. (\ref{eq:P_DtN}) to
\begin{equation}  \label{eq:P_DtN_disk_ext}
\P_{\s_0}\{ \ell_t > \ell\} = \L^{-1}_t \biggl\{\frac{1}{p} \exp\biggl(- \ell \sqrt{p/D} \frac{K_1(R \sqrt{p/D})}{K_0(R\sqrt{p/D})} \biggr) \biggr\} , 
\end{equation}
whereas the probability density $\rho(\ell,t)$ follows from
Eq. (\ref{eq:rho_P}).  The mean boundary local time is
\begin{equation}  \label{eq:mean_DtN_disk_ext}
\E\{ \ell_t \} = \L^{-1}_t \left\{ \frac{1}{p} \,  \frac{K_0(R\sqrt{p/D})}{\sqrt{p/D} \, K_1(R \sqrt{p/D})} \right\} \,.
\end{equation}
We note that Eqs. (\ref{eq:P_DtN_disk_ext},
\ref{eq:mean_DtN_disk_ext}) also characterize the boundary local time
of reflected Brownian motion outside a cylinder of radius $R$.  For
instance, $\ell_t$ describes the number of bulk relocations on a
cylindrical strand, which is relevant, e.g., in a field cycling NMR
dispersion technique \cite{Levitz08}.

The short-time behavior is the same as for the interior problem: $\E\{
\ell_t \} \simeq 2\sqrt{Dt}/\sqrt{\pi}$, in agreement with
Eq. (\ref{eq:mean_DtN1_t0}).  In turn, the long-time behavior is
different, as can be seen by looking at the limit $p\to 0$.  The
asymptotic properties of the modified Bessel functions imply that the
smallest eigenvalue $\mu_0^{(p)}$ approaches $0$ logarithmically
slowly:
\begin{equation}  \label{eq:mu0_p_disk_ext}
\mu_0^{(p)} \simeq \frac{1}{R(-\ln(R\sqrt{p/D}/2) - \gamma)}  \qquad (p\to 0) ,
\end{equation}
where $\gamma \approx 0.5772\ldots$ is the Euler constant.  As a
consequence, 
\begin{equation}  \label{eq:mean_2DE} 
\E\{ \ell_t \} \simeq R \bigl(\ln(\sqrt{4Dt}/R) - \gamma/2\bigr) + o(1) \quad (t\to\infty),
\end{equation}
i.e., the boundary local time continues to grow (in agreement with the
recurrent character of two-dimensional Brownian motion) but the growth
is logarithmically slow.

It is also instructive to determine the long-time asymptotic behavior
of the variance of $\ell_t$.  Substituting
Eq. (\ref{eq:mu0_p_disk_ext}) into Eq. (\ref{eq:mean_DtN}) with $k =
2$, one gets as $t\to\infty$:
\begin{align*}
& \E\{\ell_t^2\} 
\simeq 2R^2 \L^{-1}_t \left\{ \frac{(-\ln(R\sqrt{p/D}/2) - \gamma)^2}{p} \right\} \\
& \simeq R^2 \left\{ 2\biggl(\ln (\sqrt{4Dt}/R) - \gamma/2 \biggr)^2 - \frac{\pi^2}{12} + o(1) \right\} ,
\end{align*}
so that
\begin{equation}
\var\{\ell_t\} \simeq R^2 \left\{ \biggl(\ln (\sqrt{4Dt}/R) - \gamma/2 \biggr)^2 - \frac{\pi^2}{12} + o(1) \right\} .
\end{equation}  
The relative width of the distribution,
$\sqrt{\var\{\ell_t\}}/\E\{\ell_t\}$, slowly approaches $1$ in this
limit.

Figure \ref{fig:rho_disk}b illustrates the behavior of $\rho(\ell,t)$,
which is drastically different from the case of diffusion inside the
disk (Fig. \ref{fig:rho_disk}a).  The PDF does not have a maximum.  At
any time $t$, $\rho(\ell,t)$ exhibits a flat behavior at small $\ell$
and then drops at large $\ell$.  Moreover, the curves are getting very
close to each other at long times.  Even though this observation may
suggest an approach to a steady-state limit, this is not the case,
given that the mean boundary local time slowly grows, see
Eq. (\ref{eq:mean_2DE}).

In a similar way, one can derive the exact distribution of the
boundary local time for an annulus between two concentric circles.
Moreover, one can look for the local time on each circle or impose an
absorbing boundary condition on one of the circles.  In all these
cases, the eigenfunctions of the Dirichlet-to-Neumann operator remain
unchanged, while the eigenvalues can be written explicitly in terms of
modified Bessel functions.

\subsection{Interior of a ball}

For the ball of radius $R$, $\Omega = \{ \x\in\R^3 ~:~ |\x| < R\}$,
the eigenfunctions of the Dirichlet-to-Neumann operator are the
(normalized) spherical harmonics, $Y_{mn}(\theta,\phi)/R$ (with $n =
0,1,2,\ldots$ and $m = -n,\ldots,n$), whereas the eigenvalues are
\begin{equation}  \label{eq:mu_ballI}
\mu_n^{(p)} = \sqrt{p/D} \, \frac{i'_n(R\sqrt{p/D})}{i_n(R\sqrt{p/D})} \quad (n=0,1,2,\ldots),
\end{equation}
where $i_n(z)$ are the modified spherical Bessel functions of the
first kind.  The orthogonality of spherical harmonics to a constant
function reduces Eq. (\ref{eq:P_DtN}) to
\begin{align}  \label{eq:P_DtN_sphere}
& \P_{\s_0}\{ \ell_t > \ell\} \\  \nonumber
& = \L^{-1}_t \left\{\frac{1}{p} \exp\left(- \ell \biggl( \sqrt{p/D}\, \ctanh(R \sqrt{p/D}) - 1/R \biggr) \right)\right\} ,
\end{align}
where we used the explicit form $i_0(z) = \sinh(z)/z$.  The
probability density $\rho(\ell,t)$ follows from Eq. (\ref{eq:rho_P}).

\begin{figure}
\begin{center}
\includegraphics[width=88mm]{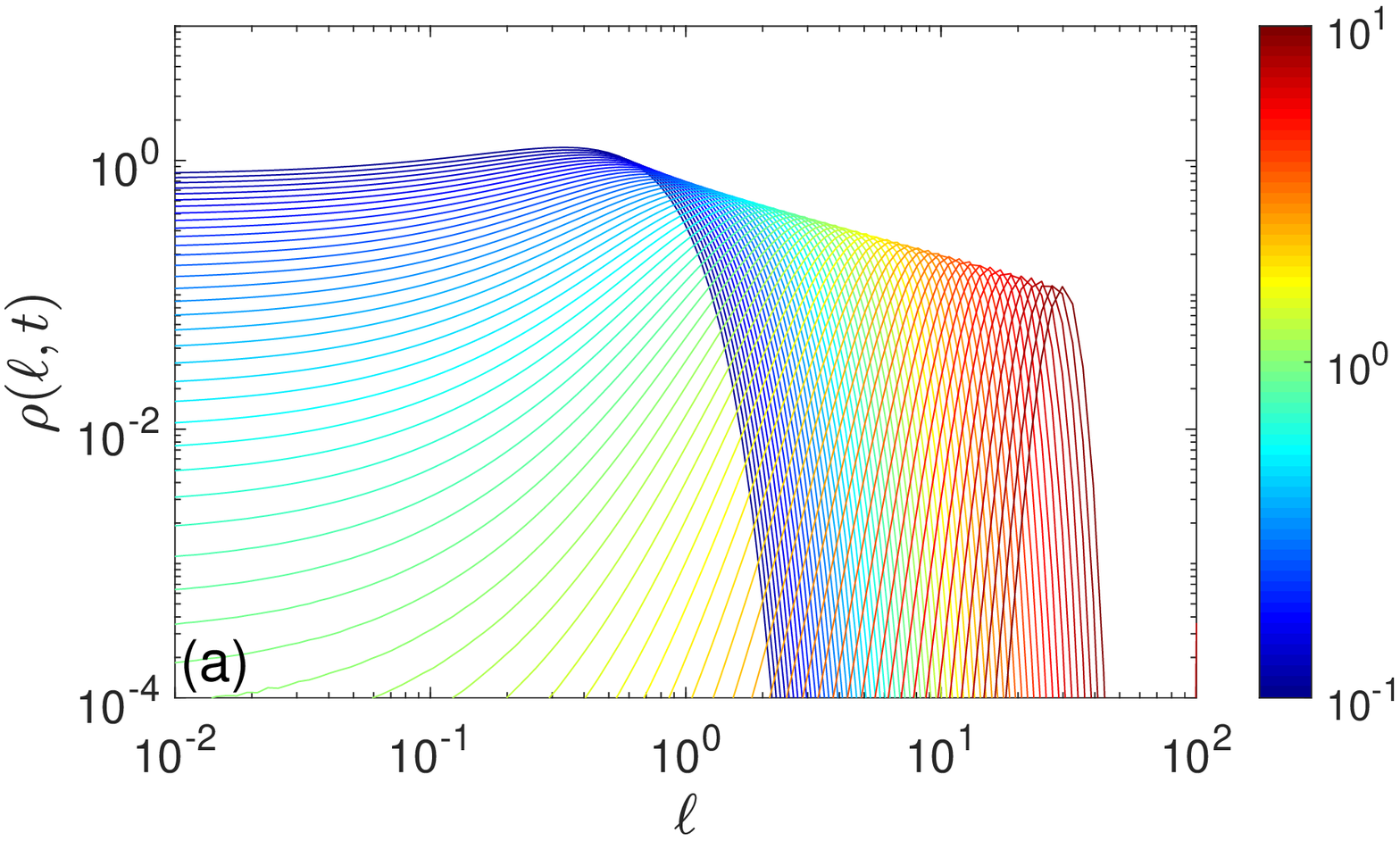} 
\includegraphics[width=88mm]{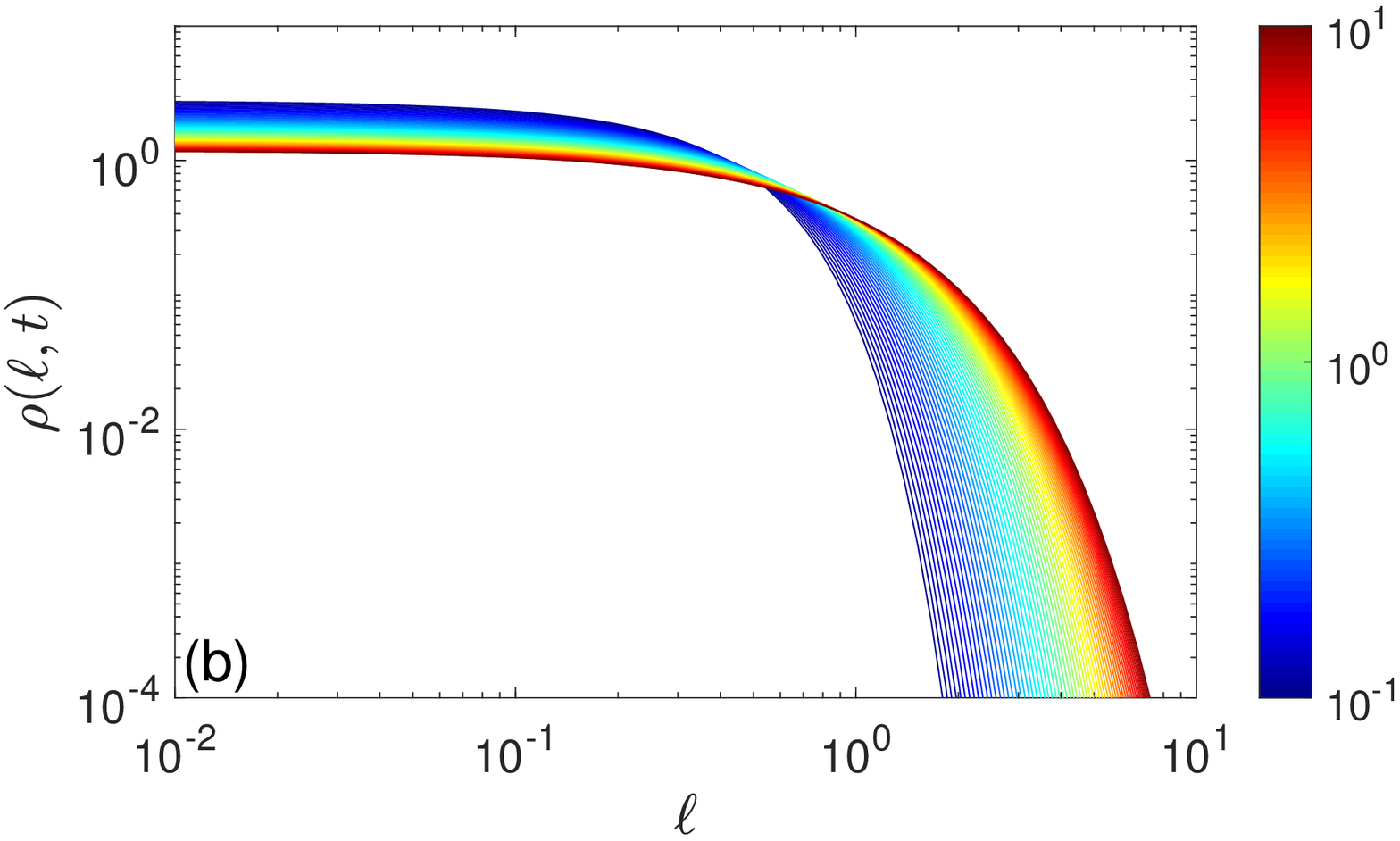} 
\end{center}
\caption{
Probability density functions $\rho(\ell,t)$ of the boundary local
time $\ell_t$ for a ball of radius $R = 1$, with $D = 1$ and $t$
taking 64 logarithmically spaced values from $10^{-1}$ (dark blue) to
$10^1$ (dark red).  {\bf (a)} diffusion inside the ball, {\bf (b)}
diffusion outside the ball.}
\label{fig:rho_ball}
\end{figure}

Figure \ref{fig:rho_ball}a illustrates the behavior of $\rho(\ell,t)$,
which is very similar to the case of diffusion inside a disk
(Fig. \ref{fig:rho_disk}a).

\subsection{Exterior of a ball}

For the exterior of a ball of radius $R$, $\Omega = \{ \x\in\R^3 ~:~
|\x|> R\}$, the eigenfunctions of the Dirichlet-to-Neumann operator
remain unchanged, whereas the eigenvalues are
\begin{equation}   \label{eq:mu_ballE}
\mu_n^{(p)} = - \sqrt{p/D} \, \frac{k'_n(R\sqrt{p/D})}{k_n(R\sqrt{p/D})} \quad (n = 0,1,2,\ldots),
\end{equation}
where $k_n(z)$ are the modified spherical Bessel functions of the
second kind.  Interestingly, the eigenvalues are just polynomials of
$\sqrt{p/D}$, e.g., $\mu_0^{(p)} = (1 + R\sqrt{p/D})/R$.  The
orthogonality of spherical harmonics implies then
\begin{align}  \nonumber
\P_{\s_0}\{ \ell_t > \ell\} & = \L^{-1}_t\left\{ \frac{1}{p} \, \exp\left(-\ell \bigl(1/R + \sqrt{p/D}) \right) \right\} \\  \label{eq:P_DtN_sphere_ext}
& = \erfc\biggl(\frac{\ell}{\sqrt{4Dt}}\biggr) \, e^{-\ell/R} \,,
\end{align}
where $\erfc(z)$ is the complementary error function.  Here, we
managed to obtain the fully explicit form of this probability.  The
probability density $\rho(\ell,t)$ follows again from
Eq. (\ref{eq:rho_P}):
\begin{equation}
\rho(\ell,t) = \frac{e^{-\ell/R}}{R} \biggl( \erfc\biggl(\frac{\ell}{\sqrt{4Dt}}\biggr) +  \frac{R \exp(-\ell^2/(4Dt))}{\sqrt{\pi Dt}}\biggr).
\end{equation}
The mean boundary local time reads
\begin{equation}
\E\{ \ell_t \} = R \bigl(1 - \erfcx(\sqrt{Dt}/R)\bigr) ,
\end{equation}
where $\erfcx(z) = e^{z^2}\erfc(z)$ is the scaled complementary error
function.  At short times, one has $\E\{ \ell_t \} \simeq
2\sqrt{Dt}/\sqrt{\pi}$, whereas at long times, $\E\{ \ell_t
\}$ approaches $R$.

Figure \ref{fig:rho_ball}b presents the behavior of $\rho(\ell,t)$.
Even though this figure looks very similar to Fig. \ref{fig:rho_disk}b
for diffusion outside a disk, there is a substantial difference: due
to the transient character of Brownian motion, the curves of
$\rho(\ell,t)$ approach their steady-state limit $\rho(\ell,\infty) =
e^{-\ell/R}/R$.  This distribution is considerably different from the
Gaussian one for diffusion in bounded domains.

In a similar way, one can derive the exact distribution of the
boundary local time for a region between two concentric spheres.
Moreover, one can look for the local time on each sphere or impose an
absorbing boundary condition on one of the spheres.  In all these
cases, the eigenfunctions of the Dirichlet-to-Neumann operator remain
unchanged, while the eigenvalues can be written explicitly in terms of
modified spherical Bessel functions.

\section{Conclusion}
\label{sec:conclusions}

In summary, we presented a general description of the boundary local
time $\ell_t$ of reflected Brownian motion in Euclidean domains.  This
description relies on the recent spectral representation of the
distribution of stopping times on partially reflecting boundaries in
terms of the Dirichlet-to-Neumann operator $\M_p$.  As stopping occurs
when $\ell_t$ exceeds a random threshold, one can access the boundary
local time as well.  The derived spectral representations
(\ref{eq:rho_DtN}, \ref{eq:P_DtN}) involve the eigenvalues and
eigenfunctions of $\M_p$ which depend only on the shape of the
confining domain.  From these general results, the short-time and
long-time asymptotic behaviors of the boundary local time were
investigated.  In particular, three geometrical settings could be
distinguished as $t\to\infty$: (i) diffusion in any bounded domain,
for which the distribution of $\ell_t$ approaches a Gaussian one, with
mean and variance growing linearly with time $t$; (ii) diffusion
outside a bounded planar set, for which the distribution is not
Gaussian and its shape varies very slowly with $t$, and (iii)
diffusion outside a bounded set in $\R^d$ with $d \geq 3$, for which
$\ell_t$ reaches a steady-state distribution.  We illustrated the
general properties of the boundary local time for five settings, for
which the spectral properties of the Dirichlet-to-Neumann operator are
known explicitly, namely, diffusion inside and outside a disk and a
ball, as well as in a half-space.  For all these cases, we derived
exact formulas for the probability density function of $\ell_t$;
moreover, in the case of diffusion outside the ball, the formulas are
fully explicit.  While the short-time asymptotic formula
(\ref{eq:mean_DtN1_t0}) for the mean boundary local time is universal,
$\E\{ \ell_t \} \propto \sqrt{t}$, the long-time behavior is not; in
fact, $\E\{ \ell_t \}$ exhibited a linear growth with $t$ for the
interior of a disk and a sphere, a logarithmical growth with $t$ for
the exterior of a disk, and an approach to a constant for the exterior
of a sphere.  This distinction reflects recurrent-versus-transient
character of Brownian motion in these domains.  In the latter case,
the steady-state value $\E\{ \ell_\infty \}$ is equal to $R$, the only
nontrivial length scale of the problem in the limit $t\to\infty$.

As discussed in \cite{Grebenkov19}, the Dirichlet-to-Neumann operator
can represent the whole propagator and thus contains equivalent
information to describe diffusion-reaction processes.  In this light,
the eigenfunctions $v_n^{(p)}(\s)$ of the Dirichlet-to-Neumann
operator $\M_p$ present an alternative to the conventional
eigenfunctions $u_n(\x)$ of the Laplace operator $\Delta_{\x}$.  The
former ones have several advantages: (i) the eigenfunctions
$v_n^{(p)}$ live on the boundary $\pa\subset \R^{d-1}$ and thus have
the reduced dimensionality as compared to the eigenfunctions $u_n$
living on $\Omega \subset \R^d$; (ii) the spectral expansions over
$v_n^{(p)}$ are available whenever the boundary is bounded, even for
unbounded domains, for which the spectrum of the Laplace operator is
continuous and thus conventional spectral expansions over $u_n$ cannot
be used; and (iii) $v_n^{(p)}$ do not depend on the reactivity
$\kappa$ of the boundary, in contrast to $u_n$.  In fact, as the
reactivity stands as the parameter of Robin boundary condition, it
enters {\it implicitly} into the propagator, the Laplacian
eigenfunctions $u_n$ and related quantities and thus remains entangled
with the shape of the domain \cite{Grebenkov13}.  In turn, the present
approach characterizes repeated returns of the particle to the
boundary via the boundary local time, which is coupled to the
reactivity {\it afterward} via the stopping time $\T$.  Here, the
shape of the domain is captured via the Dirichlet-to-Neumann operator,
while the reactivity $\kappa$ appears {\it explicitly} in spectral
expansions and is thus disentangled from the geometry.  In particular,
formula (\ref{eq:S_rho}) expresses the survival probability
$S_q(t|\x_0)$ (determining the associated first-passage time $\T$) as
the Laplace transform of the probability density of the boundary local
time.  Once the latter is known, the distribution of the first-passage
time can be accessed via this relation, for any reactivity $\kappa$.
The boundary local time is therefore the fundamental key concept in
the description of diffusion-mediated events on reactive surfaces.  As
a consequence, the current work lays the theoretical ground to better
understand the interplay between the geometrical structure of the
confining domain and its reactivity, and ultimately to control and
optimize various diffusion-reaction processes.

\appendix
\section{Asymptotic behavior of eigenvalues}
\label{sec:mu0_p}

For a bounded domain, the asymptotic behavior of the eigenvalues of
the Dirichlet-to-Neumann operator at small $p$ can be obtained via a
standard perturbation theory.  For an eigenpair $\{\mu^{(p)},
v^{(p)}\}$, one expects
\begin{align*}
v^{(p)} &= v_{(0)} + p \,v_{(1)} + O(p^2) ,  \\
\mu^{(p)} &= \mu_{(0)} + p \, \mu_{(1)} + O(p^2).
\end{align*}
Let $u^{(p)}$ denote the solution of the modified Helmholtz equation
(\ref{eq:u_def1}) with $f = v^{(p)}$ in the Dirichlet boundary
condition (\ref{eq:u_def2}).  Setting 
\begin{equation*}
u^{(p)} = u_{(0)} + p \, u_{(1)} + O(p^2)
\end{equation*}
and identifying the terms of the same order in $p$ in
Eqs. (\ref{eq:u_def}), one sees that $u_{(0)}$ and $u_{(1)}$ are
solutions of the following boundary value problems:
\begin{eqnarray}  \label{eq:auxil20}
&& D\Delta u_{(0)} = 0 \quad (\textrm{in}~\Omega), \quad u_{(0)}|_{\pa} = v_{(0)} , \\  \label{eq:auxil21}
&& D\Delta u_{(1)} = u_{(0)}  \quad (\textrm{in}~\Omega), \quad u_{(1)}|_{\pa} = v_{(1)} .
\end{eqnarray}
At the same time, the definition of the Dirichlet-to-Neumann operator
implies
\begin{align}  
(\partial_n u^{(p)})|_{\pa} & = \M_p v^{(p)} = \mu^{(p)} v^{(p)} \\  \nonumber
& = \bigl(\mu_{(0)} + p \mu_{(1)} + \ldots\bigr) \bigl(v_{(0)} + p v_{(1)} + \ldots\bigr),
\end{align}
from which the identification of the terms with the same $p$ yields
\begin{eqnarray}  \label{eq:auxil22a}
(\partial_n u_{(0)})|_{\pa} &=& \mu_{(0)} v_{(0)} , \\   \label{eq:auxil22}
(\partial_n u_{(1)})|_{\pa} &=& \mu_{(0)} v_{(1)} + \mu_{(1)} v_{(0)} .
\end{eqnarray}
According to Eqs. (\ref{eq:auxil20}, \ref{eq:auxil22a}), $\mu_{(0)}$
and $v_{(0)}$ are expectedly an eigenvalue and an eigenfunction of the
operator $\M_0$: $\M_0 v_{(0)} = \mu_{(0)} v_{(0)}$.

The solution of the boundary value problem (\ref{eq:auxil21}) can be
searched as a linear combination of two solutions: $u_{(1)} =
u_{(1)}^{\rm inh} + u_{(1)}^{\rm hom}$, with
\begin{eqnarray}
&& D\Delta u_{(1)}^{\rm inh} = u_{(0)}, \quad u_{(1)}^{\rm inh}|_{\pa} = 0 , \\
&& D\Delta u_{(1)}^{\rm hom} = 0 , \quad u_{(1)}^{\rm hom}|_{\pa} = v_{(1)} .
\end{eqnarray}
As a consequence, one can rewrite Eq. (\ref{eq:auxil22}) as
\begin{equation}
(\partial_n u_{(1)}^{\rm inh})|_{\pa} + (\partial_n u_{(1)}^{\rm hom})|_{\pa} = \mu_{(0)} v_{(1)} + \mu_{(1)} v_{(0)} .
\end{equation}
Rewriting the second term on the left-hand side as $\M_0 v_{(1)}$,
multiplying this relation by $v_{(0)}$ and integrating over $\pa$, one
gets
\begin{equation}
\bigl(v_{(0)} \cdot \partial_n u_{(1)}^{\rm inh}\bigr)_{L_2(\pa)}  = \mu_{(1)} ,
\end{equation}
where we used the $L_2(\pa)$-normalization of $v_{(0)}$ as an
eigenfunction of $\M_0$, and $(v_{(0)} \cdot \M_0 v_{(1)})_{L_2(\pa)}
= \mu_{(0)} (v_{(0)} \cdot v_{(1)})_{L_2(\pa)}$ because $\M_0$ is
self-adjoint.

For the lowest eigenpair, with $\mu_{(0)} = 0$ and $v_{(0)} =
|\pa|^{-1/2}$, one gets
\begin{align}  \nonumber
\mu_{(1)} &= |\pa|^{-1/2} \int\limits_{\pa} d\s \, \partial_n u_{(1)}^{\rm inh}  \\
& = |\pa|^{-1/2} \int\limits_{\Omega} d\x \, \underbrace{\Delta u_{(1)}^{\rm inh}}_{= u_{(0)}}
= \frac{|\Omega|}{D|\pa|} \,,
\end{align}
where we used that $u_{(0)}$ is a constant solution of
Eq. (\ref{eq:auxil20}) subject to the constant boundary condition
$v_{(0)} = |\Omega|^{-1/2}$.  We conclude that
\begin{equation}
\mu_0^{(p)} \simeq \frac{|\Omega|}{D|\pa|}\,  p + O(p^2) \qquad (p\to 0) \,.
\end{equation}

\section{Variance of the boundary local time}
\label{sec:b21}

In \cite{Grebenkov07a}, the long-time asymptotic behavior of the
cumulant moments of the residence time and other functionals of
reflected Brownian motion was investigated.  In particular, the
variance of $\ell_t$ was shown to be
\begin{equation}
\var\{\ell_t\} \simeq b_{2,1} t + b_{2,0}  \qquad (t\to\infty),
\end{equation}
with two constants $b_{2,1}$ and $b_{2,0}$ depending on the domain
$\Omega$.  For a bounded domain, the constant of the leading term
reads
\begin{equation}  \label{eq:b21_lambda}
b_{2,1} = \frac{2}{D} \sum\limits_{m=1}^\infty \lambda_m^{-1} B_{0,m}^2 ,
\end{equation}
where $\lambda_m$ (with $m=0,1,2,\ldots$) are the eigenvalues of the
Laplace operator in $\Omega$ with Neumann boundary condition on $\pa$, and 
\begin{equation}
B_{m,m'} = \int\limits_\Omega d\x \, u_m^*(\x) \, B(\x) \, u_{m'}(\x),
\end{equation}
where $u_m(\x)$ are the corresponding eigenfunctions of the Laplace
operator, and $B(\x)$ is the considered functional.  Note that the
ground eigenmode with $m = 0$ (corresponding to $\lambda_0 = 0$ and
$u_0 = |\Omega|^{-1/2}$) is excluded from the sum in
Eq. (\ref{eq:b21_lambda}).

In the case of the boundary local time, Eq. (\ref{eq:ell_res}) implies
that $B(\x)$ is proportional to the indicator function of the vicinity
$\pa_a$ of the boundary: $B(\x) = \frac{D}{a} \I_{\pa_a}(\x)$.  Taking
the limit $a\to 0$, one gets:
\begin{equation}
B_{m,m'} = D \int\limits_{\pa} d\s \, u_m^*(\s) \, u_{m'}(\s).
\end{equation}
As a consequence, the constant $b_{2,1}$ can be written as
\begin{equation}
b_{2,1} = \frac{2D}{|\Omega|} \int\limits_{\pa} d\s_1 \int\limits_{\pa} d\s_2 \, \sum\limits_{m=1}^\infty u_m^*(\s_1) \, u_m(\s_2) \lambda_m^{-1} .
\end{equation}
Writing the Laplace-transformed propagator as
\begin{equation}
\tilde{G}_{0}(\s,p|\s_0) = \sum\limits_{m=0}^\infty \frac{u_m^*(\s) \, u_m(\s_0)}{p + D\lambda_m} \,,
\end{equation}
we subtract the ground mode with $m = 0$ to get
\begin{equation}
b_{2,1} = \frac{2D}{|\Omega|} \int\limits_{\pa} d\s_1 \int\limits_{\pa} d\s_2 \, {\mathcal G}(\s_1,\s_2) ,
\end{equation}
where
\begin{equation}
{\mathcal G}(\s,\s_0) = D \lim\limits_{p\to 0}  \biggl(\tilde{G}_{0}(\s,p|\s_0) - \frac{1}{p|\Omega|}\biggr)
\end{equation}
is the pseudo-Green function.  The subtraction of the ground mode,
which diverges in the limit $p\to 0$, can be seen a regularization of
the Laplace-transformed propagator.  In fact,
$\tilde{G}_{0}(\s,p|\s_0)$ diverges as $p\to 0$, in agreement with the
well-known statement that the Green function of the Laplace operator
(i.e., for $p = 0$) in a bounded domain with Neumann boundary
condition does not exist.
Using the fact that $D \tilde{G}_{0}(\s,p|\s_0)$ is the kernel of
$\M_p^{-1}$ due to Eq. (\ref{eq:Mp_inv}), we get
\begin{equation}
b_{2,1} = \frac{2D}{|\Omega|} \lim\limits_{p\to 0} \biggl(\bigl(1 , \M_p^{-1} 1 \bigr)_{L_2(\pa)} - \frac{D |\pa|^2}{p|\Omega|} \biggr).
\end{equation}
Finally, expanding the above scalar product on the eigenbasis of
$\M_p$, one has
\begin{align}  \nonumber
b_{2,1} & = \frac{2D}{|\Omega|} \lim\limits_{p\to 0} \biggl(\frac{|(v_0^{(p)} , 1)_{L_2(\pa)}|^2}{\mu_0^{(p)}}
- \frac{D |\pa|^2}{p|\Omega|} \\
& + \sum\limits_{n=1}^\infty \frac{|(v_n^{(p)} , 1)_{L_2(\pa)}|^2}{\mu_n^{(p)}} \biggr) ,
\end{align}
where we wrote separately the term with $n = 0$.  In the limit $p\to
0$, the eigenfunctions $v_n^{(p)}$ tend to $v_n^{(0)}$, which are
orthogonal to $v_0^{(0)} = |\pa|^{-1/2}$.  As a consequence, the last
term vanishes in this limit, and we are left with
\begin{equation}
b_{2,1} = \frac{2D|\pa|}{|\Omega|} \lim\limits_{p\to 0} \biggl(\frac{1}{\mu_0^{(p)}}
- \frac{D |\pa|}{p|\Omega|} \biggr).
\end{equation}
Expanding the smallest eigenvalue $\mu_0^{(p)}$ into a series in
powers of $p$, $\mu_0^{(p)} = 0 + p \mu_{(1)} + \frac12 p^2 \mu_{(2)}
+ \ldots$, one finally gets
\begin{equation}
b_{2,1} = - \biggl(\frac{D |\pa|}{|\Omega|}\biggr)^3 \lim\limits_{p\to 0} \frac{d^2 \mu_0^{(p)}}{dp^2} \,.
\end{equation}
Interestingly, while the first derivative of $\mu_0^{(p)}$ at $p = 0$
determines the asymptotic mean of the boundary local time, the second
derivative determines its variance.

\section{Validation by Monte Carlo simulations} 
\label{sec:validation}

In order to validate our analytical results and the quality of the
numerical Laplace transform inversion, we undertake Monte Carlo
simulations of reflected Brownian motion with diffusion coefficient
$D$ inside a disk and a ball of radius $R$.  We employ a basic fixed
time-step scheme, even though more advanced Monte Carlo techniques are
available \cite{Morillon97,Costantini98,Grebenkov14a,Zhou16,Zhou17}.
We set $R = 1$ and $D = 1$ to fix units of length and time.  For a
fixed time step $\delta$, each jump is generated independently as a
Gaussian displacement with mean zero and variance $2D\delta$ in each
spatial direction.  When the next generated position $\x$ appears
outside the domain, it is replaced by a reflected position $\x' = \x
(2R - |\x|)/|\x|$ inside the domain, which is at the same distance
from the boundary as $\x$.  For each simulated trajectory, we count
how long it remained in a boundary layer of width $a$ until time $t$.
If $N_t$ is the (random) number of positions of the trajectory inside
this layer, then $N_t \delta$ is a discrete approximation of the
residence time in this layer, whereas $D N_t
\delta/a$ is an approximation of the boundary local time $\ell_t$.
Simulating a large number $M$ of such trajectories, we get the
statistics of $\ell_t$ at different times $t$.  The normalized
histogram of this statistics approximates the probability density
function $\rho(\ell,t)$ of $\ell_t$.  The starting point was fixed on
the boundary (its actual location on the boundary does not matter due
to the rotation symmetry).

The quality of Monte Carlo simulations depends on the choice of the
numerical parameters $M$, $\delta$, and $a$.  We set $M = 10^5$ to
have a good enough statistics of random realizations of $\ell_t$.  To
ensure an accurate simulation of reflected Brownian motion, the
typical size of individual jumps, $\sqrt{2D\delta}$, should be the
smallest length scale, i.e., $\sqrt{2D\delta} \ll a$.  We fix $\delta
= 10^{-5}$ to get $\sqrt{2D\delta} \approx 0.0045$.  To check the
consistence of simulated results, we performed simulations for ten
equally spaced values of $a$, from $a = 0.005$ to $a = 0.05$.  On one
hand, smaller $a$ ensures better approximation of the boundary local
time by the residence time in Eq. (\ref{eq:ell_res}).  On the other
hand, $a$ should not become smaller than $\sqrt{2D\delta}$.

\begin{figure}
\begin{center}
\includegraphics[width=88mm]{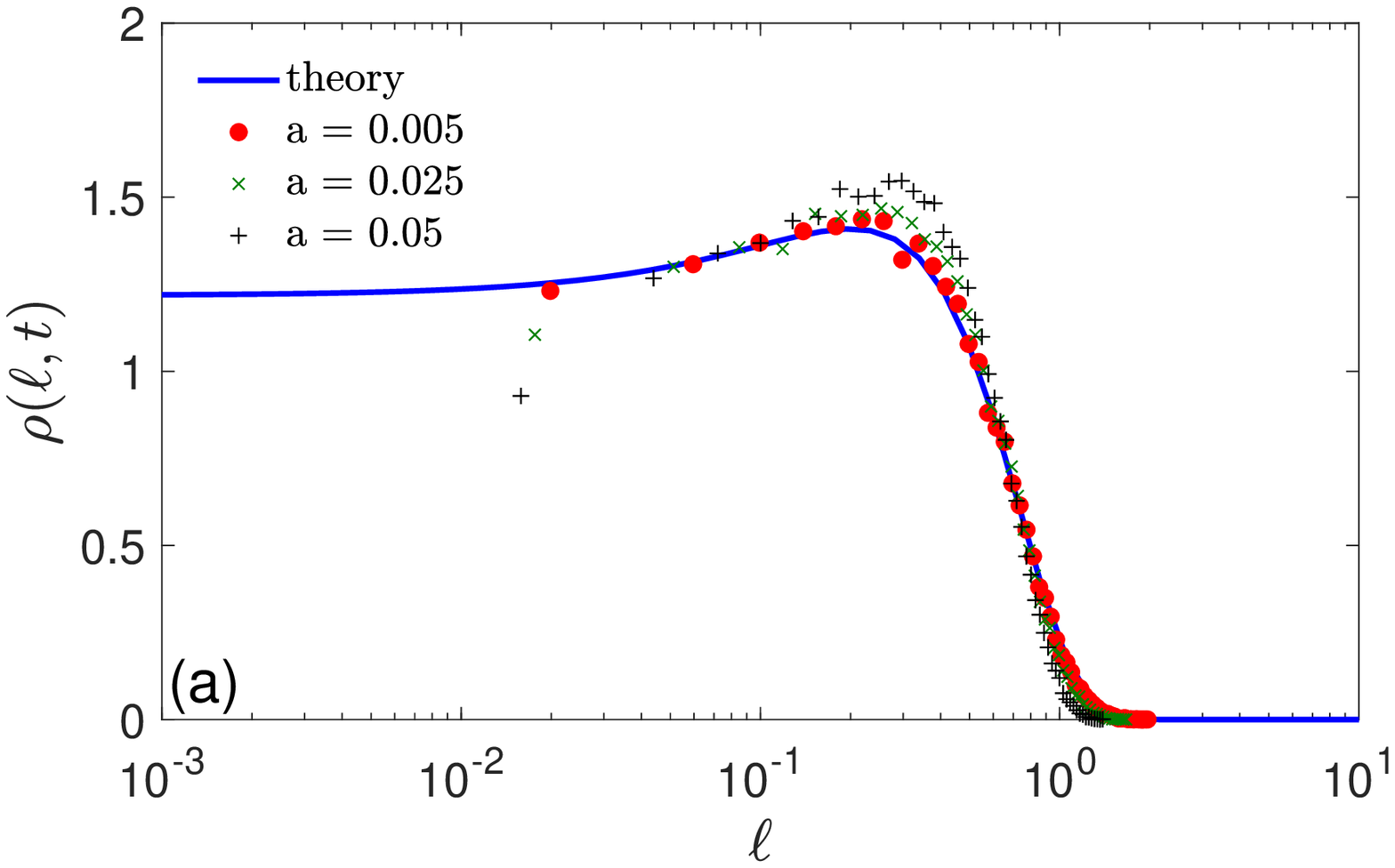} 
\includegraphics[width=88mm]{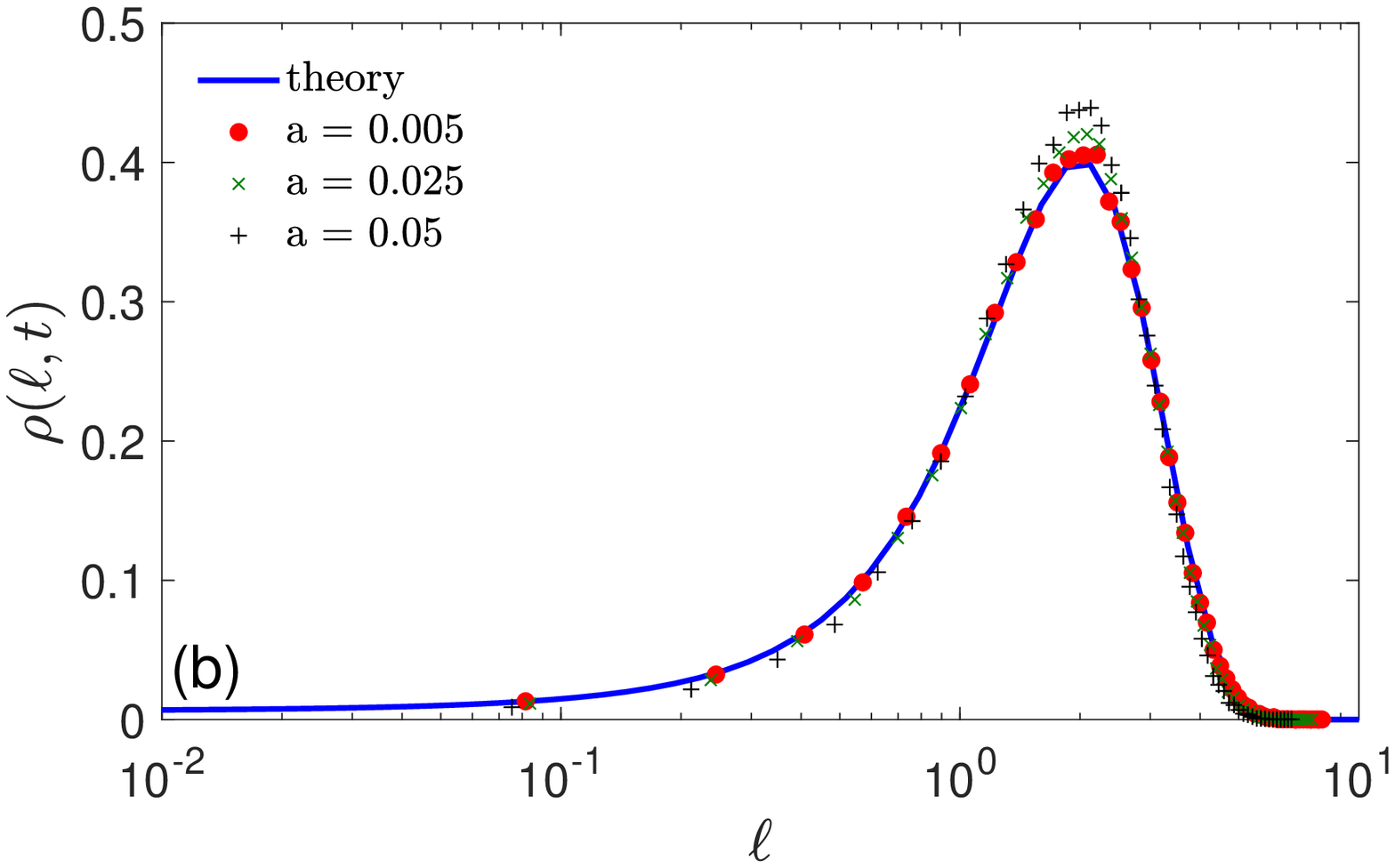} 
\includegraphics[width=88mm]{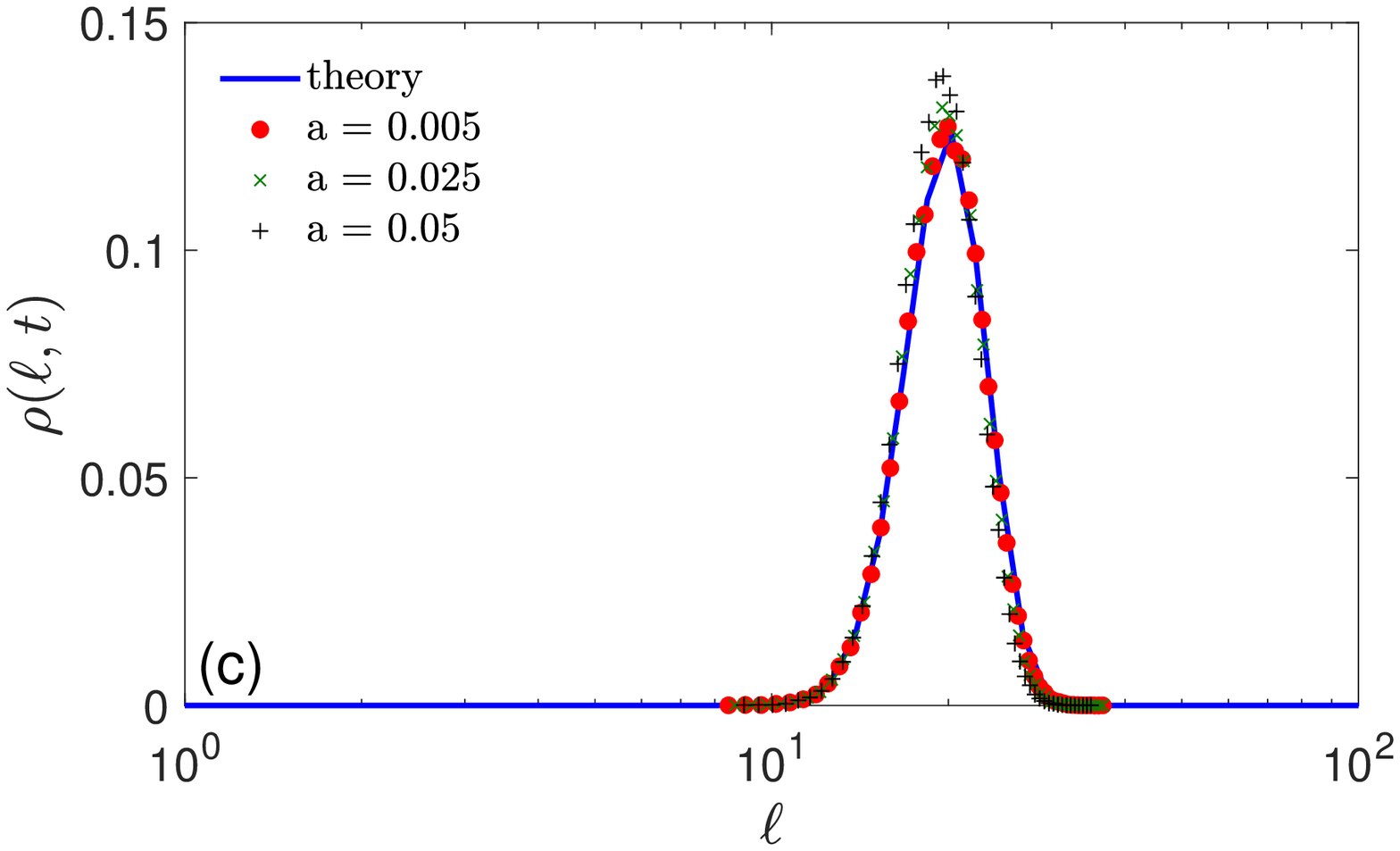} 
\end{center}
\caption{
Probability density function $\rho(\ell,t)$ of the boundary local time
$\ell_t$ for a disk of radius $R = 1$, with $D = 1$ and three values
of time: {\bf (a)} $t = 0.1$, {\bf (b)} $t = 1$, and {\bf (c)} $t =
10$.  Solid line shows numerical inversion of the Laplace transform in
Eq. (\ref{eq:rho_Laplace}), whereas symbols illustrate normalized
histograms obtained from Monte Carlo simulations, with $M = 10^5$,
$\delta = 10^{-5}$, and three values of $a$ as indicated in the
legend. }
\label{fig:rho_ell_disk}
%
\end{figure}

\begin{figure}
\begin{center}
\includegraphics[width=88mm]{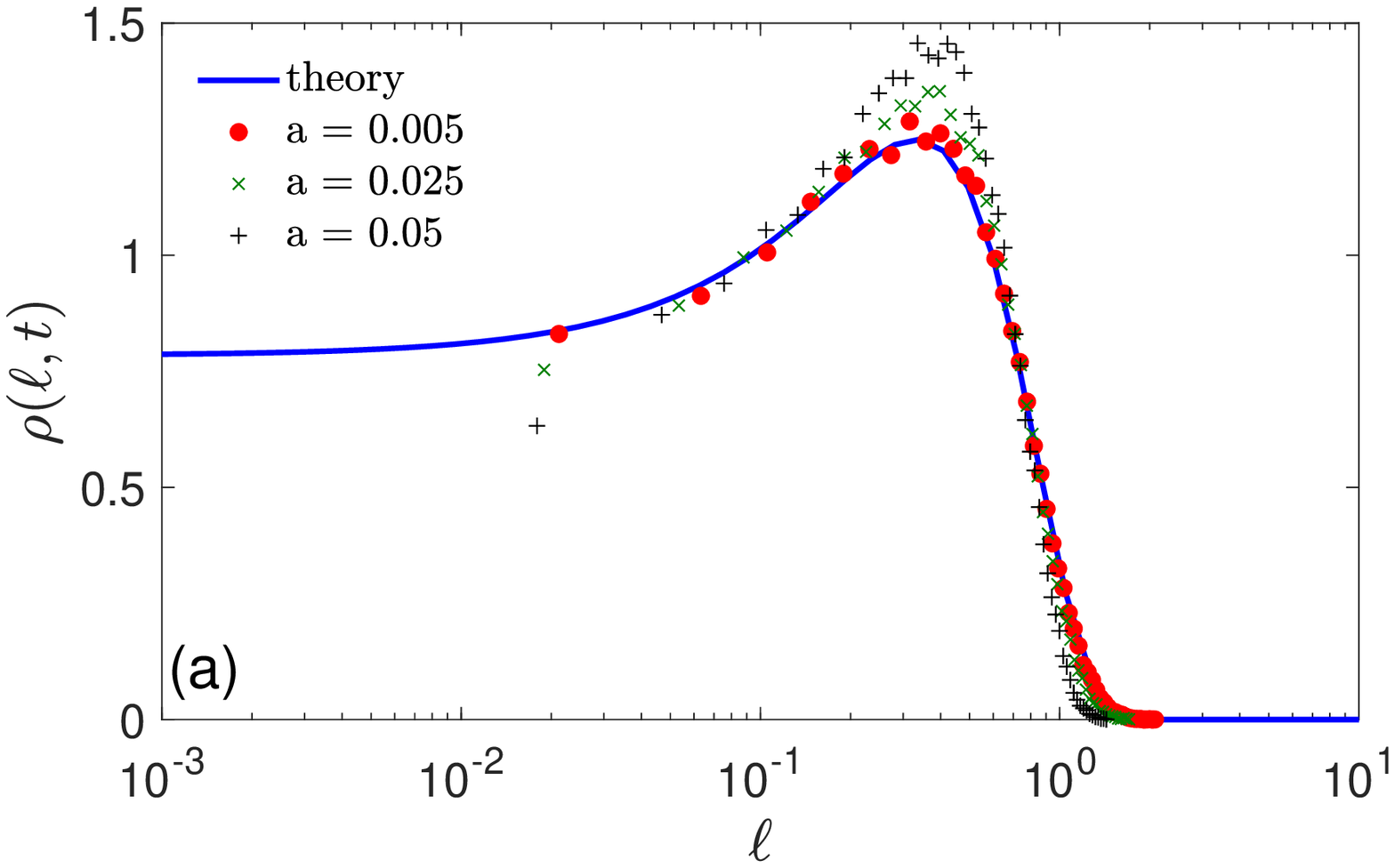} 
\includegraphics[width=88mm]{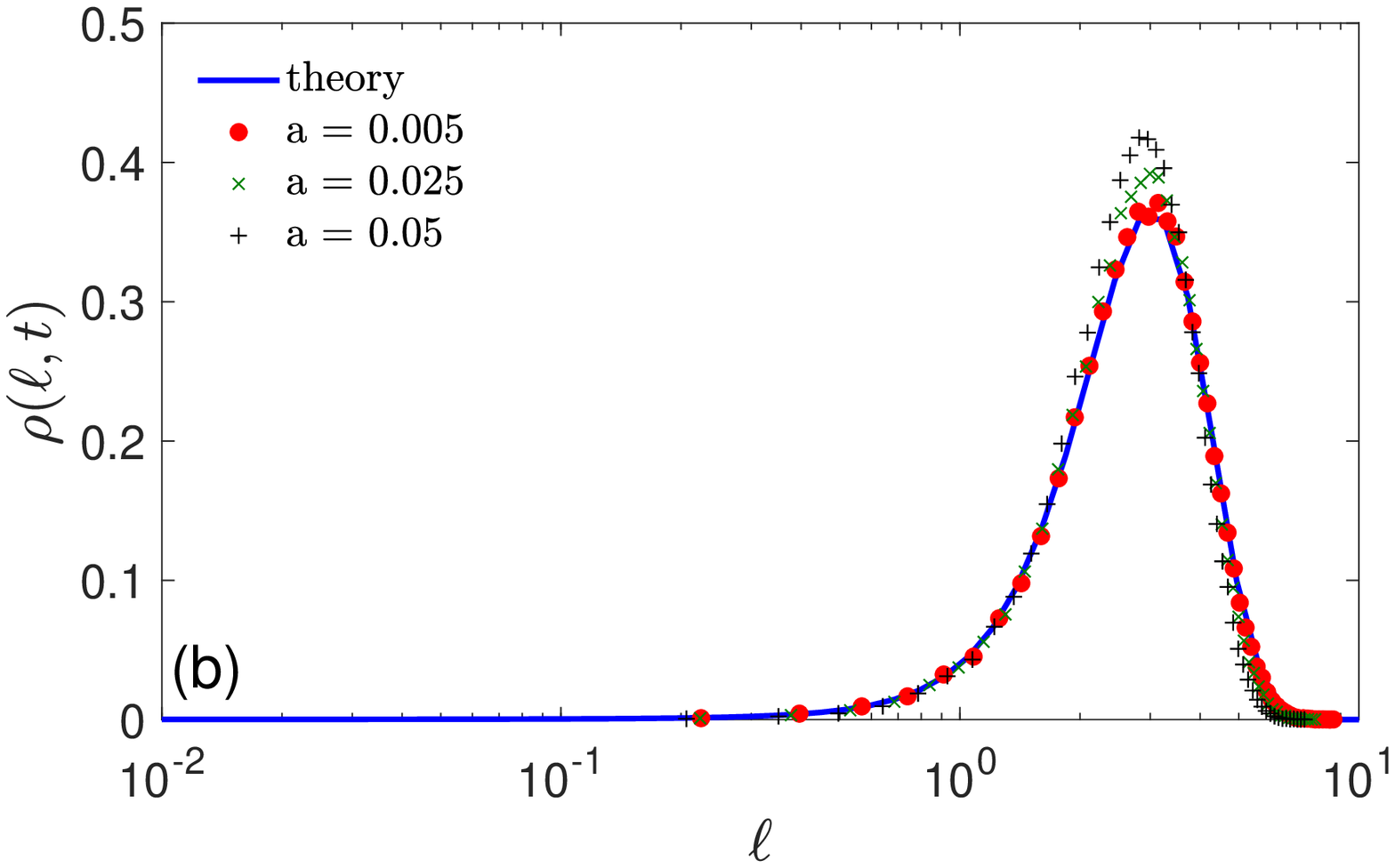} 
\includegraphics[width=88mm]{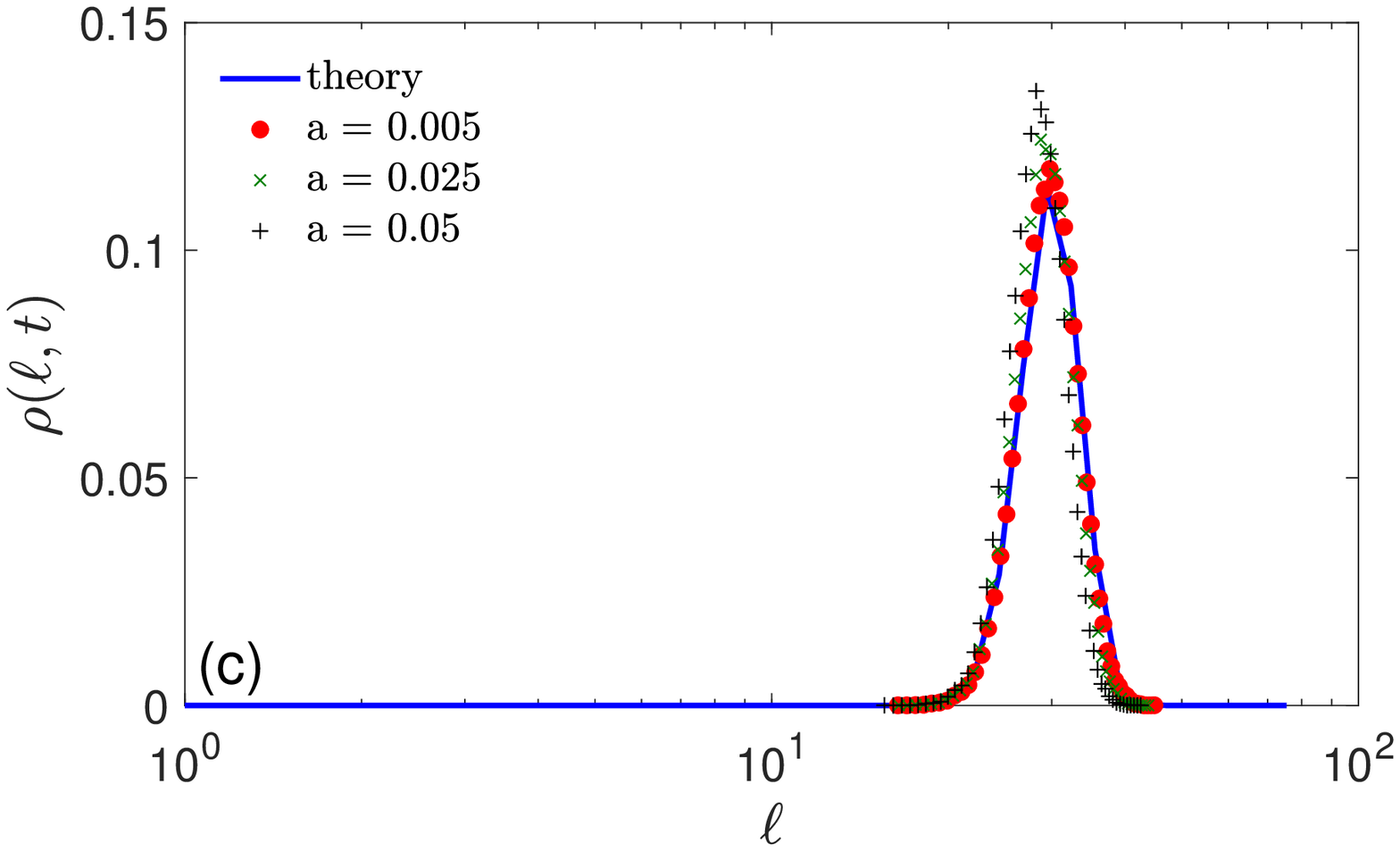} 
\end{center}
\caption{
Probability density function $\rho(\ell,t)$ of the boundary local time
$\ell_t$ for a ball of radius $R = 1$, with $D = 1$ and three values
of time: {\bf (a)} $t = 0.1$. {\bf (b)} $t = 1$, and {\bf (c)} $t =
10$.  Solid line shows numerical inversion of the Laplace transform in
Eq. (\ref{eq:rho_Laplace}), whereas symbols illustrate normalized
histograms obtained from Monte Carlo simulations, with $M = 10^5$,
$\delta = 10^{-5}$, and three values of $a$ indicated in the legend.}
\label{fig:rho_ell_ball}
%
\end{figure}

Figure \ref{fig:rho_ell_disk} shows the probability density function
$\rho(\ell,t)$ for a disk at three values of time: $t = 0.1$, $t = 1$,
and $t = 10$.  Solid line presents $\rho(\ell,t)$ evaluated via the
numerical inversion of the Laplace transform (by Talbot algorithm) in
Eq. (\ref{eq:rho_DtN_inv}), which can be written more explicitly as
\begin{equation}  \label{eq:rho_Laplace}
\rho(\ell,t) = \L^{-1}_t \left\{ \frac{\mu_0^{(p)}}{p} \exp(-\ell \mu_0^{(p)}) \right\} ,
\end{equation}
with $\mu_0^{(p)}$ given by Eq. (\ref{eq:mu_diskI}) for the disk and
by Eq. (\ref{eq:mu_ballI}) for the ball.  In turn, symbols present
$\rho(\ell,t)$ from Monte Carlo simulations for three values of $a$.
As the value of $a$ decreases, the simulated normalized histograms are
getting closer to our theoretical results, as expected.  The best
agreement is observed for $a = 0.005$, which is actually comparable to
$\sqrt{2D\delta}$.  We performed another set of simulations with
$\delta = 10^{-6}$ and thus much smaller $\sqrt{2D\delta}$, and the
obtained histograms were very close to those on
Fig. \ref{fig:rho_ell_disk} (for this reason, these histograms are not
shown).  The perfect agreement between Monte Carlo simulations and
theoretical curves can be seen as a cross-validation of simulations,
theory, and the used numerical inversion of the Laplace transform.
Figure \ref{fig:rho_ell_ball} presents very similar results for the
case of a ball.

\end{document}